\documentclass[sigconf,authorversion]{acmart}

\usepackage{xspace}
\newcommand{\x}{\texttimes\xspace}



\newcommand{\totsize}{185}
\newcommand{\morearea}{16}
\newcommand{\morepower}{12}
\newcommand{\fastercpu}{18}
\newcommand{\fastergpu}{16}

\newcommand{\arbslow}{27}
\newcommand{\arbslowmax}{90}

\newcommand{\slowvsideal}{8}
\newcommand{\cpuminfaster}{4.4}
\newcommand{\cpumaxfaster}{327}
\newcommand{\gpuminfaster}{4.9}
\newcommand{\gpumaxfaster}{118}

\usepackage{algorithmicx}
\usepackage{algpseudocode}
\usepackage{booktabs}
\usepackage{multirow}
\usepackage{multicol}
\usepackage{outlines}
\usepackage{pifont}
\usepackage{siunitx}
\usepackage{tabu}
\usepackage{todonotes}
\usepackage{xparse}
\usepackage{xspace}
\usepackage{xcolor}
\usepackage{xcolor-solarized}
\usepackage{pgfplots}
\usepackage{pgfplotstable}
\usepackage{listings}
\usepackage[T1]{fontenc}
\usepackage{enumitem}
\usepackage{textcomp}
\usepackage[caption=false, font=footnotesize]{subfig}
\usepackage{cleveref}
\newcommand{\added}[1]{#1}
\usepackage{etoolbox}
\newrobustcmd{\ubold}{\fontseries{b}\selectfont}

  \newcolumntype{R}[2]{%
    >{\adjustbox{angle=#1,lap=\width-(#2)}\bgroup}%
    l%
    <{\egroup}%
}

\usepgfplotslibrary{colorbrewer}
\usepgfplotslibrary{groupplots}
\usepgflibrary{patterns}

\colorlet{circgray}{black!70}
\newcommand{\circone}{{\color{circgray}\ding{202}}\xspace}
\newcommand{\circtwo}{{\color{circgray}\ding{203}}\xspace}
\newcommand{\circthree}{{\color{circgray}\ding{204}}\xspace}
\newcommand{\circfour}{{\color{circgray}\ding{205}}\xspace}
\newcommand{\circfive}{{\color{circgray}\ding{206}}\xspace}
\newcommand{\circsix}{{\color{circgray}\ding{207}}\xspace}
\newcommand{\circseven}{{\color{circgray}\ding{208}}\xspace}

\usetikzlibrary{calc}
\usetikzlibrary{colorbrewer}
\usetikzlibrary{positioning}
\usetikzlibrary{matrix}
\usetikzlibrary{shapes}
\usetikzlibrary{patterns}
\colorlet{codebg}{black!03}

\tikzset{every picture/.style={/utils/exec={\footnotesize\sffamily}}}

\tikzstyle{hairline}=[very thin,draw=gray]
\tikzstyle{line}=[semithick]
\tikzstyle{arrow}=[line,->,>=latex]
\tikzstyle{barplot}=[ybar]
\tikzstyle{box}=[draw=black]
\tikzstyle{textnode}=[align=center,font=\footnotesize]

\pgfplotsset{compat=newest}

\pgfplotsset{
  groupbase/.style={
    axis x line*=bottom,
    axis y line*=left,
    width=0.320\linewidth,
    group style={group size=3 by 1, horizontal sep=48pt}, 
  },
  globplot/.style={
    major grid style={draw=Greys-6-3},
    axis x line* = bottom,
    axis y line* = left,
    axis line style ={-},
    width=3.2in,
    legend cell align={left},
    legend style={
      draw=none
    },
    enlarge x limits = 0.1,
  },
  nonlog/.style={
    yticklabel={$\mathsf{\pgfmathprintnumber{\tick}}$},
  },
  logplot/.style={
  },
  barplot/.style={
    enlarge x limits = 0.2,
    globplot,
    ybar=0pt,
    ytick style={draw=none},
    cycle list name=bw_cycle5,
    bar width=9pt,
    ymajorgrids,
	/pgfplots area legend/.style={%
/pgfplots/legend image code/.code={%
\fill[##1] (0cm,-0.1cm) rectangle (0.6cm,0.1cm);
}%
},
    legend style={
	    at={(0.5,-0.6)}, 
	    anchor=north, 
	    legend columns=-1, 
	    /tikz/every even column/.append style={column sep=0.5cm}
	    },
  },
  scanheight/.style={
      height=1.15in
    },
  speedup/.style={
    barplot,
      height=0.8in,
    bar width=8pt,
enlarge x limits=0.2,
      ymin=0.5,
      xticklabel style={anchor=north east},
      ylabel=Speedup,
      cycle list shift=1,
every x tick label/.append style={font=\footnotesize, rotate=20},
    legend style={
	    at={(0.5,1.0)}, 
anchor=south
},
      xtick=data
  },
  globalbar/.style={
    set layers,
    font=\footnotesize\sffamily,
    legend style={
      font=\footnotesize\sffamily\strut,
      /tikz/every even column/.append style={column sep=0.5pc},
      draw=none
    },
    minor grid style={line width=.2pt, draw=gray!15},
    major grid style={line width=.2pt, draw=gray!70},
    ytick style={draw=none},
    every axis plot/.append style={fill, on layer=pre main},
    yticklabel={$\mathsf{\pgfmathprintnumber{\tick}}$},
    area legend,
  },
  speedup_style/.style={
    width=0.99\columnwidth,
    height=1.25in,
    ybar=0pt,
    xtick=data,
    y axis line style={draw=none},
    legend style={
      legend columns=2,
      at={(0.5,-0.3)},
      anchor=north,
    },
    bar width=1.5mm,
    enlarge x limits=0.2,
    cycle list/Blues-7,
    cycle list shift=2,
    x tick label style={text width=2cm, align=center},
    axis x line*=bottom,
    globalbar,
    ymajorgrids=true,
    yminorgrids=true,
    ytick pos=left,
  }
}

\definecolor{Greys-3-1}{RGB}{240,240,240}
\definecolor{Greys-3-C}{RGB}{240,240,240}
\definecolor{Greys-3-2}{RGB}{189,189,189}
\definecolor{Greys-3-F}{RGB}{189,189,189}
\definecolor{Greys-3-3}{RGB}{99,99,99}
\definecolor{Greys-3-I}{RGB}{99,99,99}
\definecolor{Greys-4-1}{RGB}{247,247,247}
\definecolor{Greys-4-B}{RGB}{247,247,247}
\definecolor{Greys-4-2}{RGB}{204,204,204}
\definecolor{Greys-4-E}{RGB}{204,204,204}
\definecolor{Greys-4-3}{RGB}{150,150,150}
\definecolor{Greys-4-G}{RGB}{150,150,150}
\definecolor{Greys-4-4}{RGB}{82,82,82}
\definecolor{Greys-4-J}{RGB}{82,82,82}
\definecolor{Greys-5-1}{RGB}{247,247,247}
\definecolor{Greys-5-B}{RGB}{247,247,247}
\definecolor{Greys-5-2}{RGB}{204,204,204}
\definecolor{Greys-5-E}{RGB}{204,204,204}
\definecolor{Greys-5-3}{RGB}{150,150,150}
\definecolor{Greys-5-G}{RGB}{150,150,150}
\definecolor{Greys-5-4}{RGB}{99,99,99}
\definecolor{Greys-5-I}{RGB}{99,99,99}
\definecolor{Greys-5-5}{RGB}{37,37,37}
\definecolor{Greys-5-K}{RGB}{37,37,37}
\definecolor{Greys-6-1}{RGB}{247,247,247}
\definecolor{Greys-6-B}{RGB}{247,247,247}
\definecolor{Greys-6-2}{RGB}{217,217,217}
\definecolor{Greys-6-D}{RGB}{217,217,217}
\definecolor{Greys-6-3}{RGB}{189,189,189}
\definecolor{Greys-6-F}{RGB}{189,189,189}
\definecolor{Greys-6-4}{RGB}{150,150,150}
\definecolor{Greys-6-G}{RGB}{150,150,150}
\definecolor{Greys-6-5}{RGB}{99,99,99}
\definecolor{Greys-6-I}{RGB}{99,99,99}
\definecolor{Greys-6-6}{RGB}{37,37,37}
\definecolor{Greys-6-K}{RGB}{37,37,37}

\pgfplotscreateplotcyclelist{bw_cycle3}{%
  {fill=Greys-3-1,draw=Greys-4-4},
  {fill=Greys-3-2,draw=Greys-4-4},
  {fill=Greys-3-3,draw=Greys-4-4},
}

\pgfplotscreateplotcyclelist{bw_cycle4}{%
  {fill=Greys-4-1,draw=Greys-5-5},
  {fill=Greys-4-2,draw=Greys-5-5},
  {fill=Greys-4-3,draw=Greys-5-5},
  {fill=Greys-4-4,draw=Greys-5-5},
}

\pgfplotscreateplotcyclelist{bw_cycle5}{%
  {fill=Greys-5-1,draw=Greys-6-6},
  {fill=Greys-5-2,draw=Greys-6-6},
  {fill=Greys-5-3,draw=Greys-6-6},
  {fill=Greys-5-4,draw=Greys-6-6},
  {fill=Greys-5-5,draw=Greys-6-6},
}

\pgfplotscreateplotcyclelist{solarized}{%
  {fill=solarized-base0},
  {fill=solarized-base02},
  {fill=solarized-base1},
  {fill=solarized-orange},
  {fill=solarized-blue},
  {fill=solarized-yellow},
  {fill=solarized-base03},
  {fill=solarized-red},
  {fill=solarized-cyan},
  {fill=solarized-orange},
}

\pgfplotscreateplotcyclelist{solarizedlinesapps}{%
  {draw=solarized-base03,mark=*,style=dashed},
  {draw=solarized-base03,mark=triangle*,style=dashed},
  {draw=solarized-base03,mark=square*,style=dashed},
  {draw=solarized-red,mark=pentagon*},
  {draw=solarized-cyan,style=},
  {draw=solarized-cyan,style=},
  {draw=solarized-base02,style=dotted},
  {draw=solarized-base02,style=dotted},
  {draw=solarized-orange},
  {draw=solarized-orange},
}

\pgfplotstableset{
    normpoint/.style={
        postproc cell content/.append code={
            \pgfkeysalso{@cell content/.add={\ubold}{}}
        },
    },
}

\tikzset{
        hatch distance/.store in=\hatchdistance,
        hatch distance=10pt,
        hatch thickness/.store in=\hatchthickness,
        hatch thickness=2pt
    }

    \makeatletter
    \pgfdeclarepatternformonly[\hatchdistance,\hatchthickness]{flexible hatch}
    {\pgfqpoint{0pt}{0pt}}
    {\pgfqpoint{\hatchdistance}{\hatchdistance}}
    {\pgfpoint{\hatchdistance-1pt}{\hatchdistance-1pt}}%
    {
        \pgfsetcolor{\tikz@pattern@color}
        \pgfsetlinewidth{\hatchthickness}
        \pgfpathmoveto{\pgfqpoint{0pt}{0pt}}
        \pgfpathlineto{\pgfqpoint{\hatchdistance}{\hatchdistance}}
        \pgfusepath{stroke}
    }
    \makeatother

\pgfplotscreateplotcyclelist{stallcycle}{%
      {fill=Paired-A},
      {fill=Paired-B},
      {fill=Paired-C},
      {fill=Paired-D},
      {fill=Paired-E},
      {fill=Paired-F},
      {fill=Paired-G},
      {fill=Paired-H}
    }

\newcommand{\name}{Capstan\xspace}
\newcommand{\nameupper}{CAPSTAN\xspace}

\title{\name: A Vector RDA for Sparsity}

\copyrightyear{2021}
\acmYear{2021}
\setcopyright{acmlicensed}\acmConference[MICRO '21]{MICRO'21: 54th Annual IEEE/ACM International Symposium on Microarchitecture}{October 18--22, 2021}{Virtual Event, Greece}
\acmBooktitle{MICRO'21: 54th Annual IEEE/ACM International Symposium on Microarchitecture (MICRO '21), October 18--22, 2021, Virtual Event, Greece}
\acmPrice{15.00}
\acmDOI{10.1145/3466752.3480047}
\acmISBN{978-1-4503-8557-2/21/10}

\author{Alexander Rucker}
\affiliation{%
  \institution{Stanford University}
  \city{Stanford}
  \state{CA}
  \country{USA}
}
\author{Matthew Vilim}
\affiliation{%
  \institution{Stanford University}
  \city{Stanford}
  \state{CA}
  \country{USA}
}
\author{Tian Zhao}
\affiliation{%
  \institution{Stanford University}
  \city{Stanford}
  \state{CA}
  \country{USA}
}
\author{Yaqi Zhang}
\affiliation{%
  \institution{Stanford University}
  \city{Stanford}
  \state{CA}
  \country{USA}
}
\author{Raghu Prabhakar}
\affiliation{%
  \institution{SambaNova Systems}
  \city{Palo Alto}
  \state{CA}
  \country{USA}
}
\author{Kunle Olukotun}
\affiliation{%
  \institution{Stanford University}
  \city{Stanford}
  \state{CA}
  \country{USA}
}

\begin{document}

\thispagestyle{empty}
\begin{abstract}
  This paper proposes \name: a scalable, parallel-patterns-based, reconfigurable dataflow accelerator (RDA) for sparse and dense tensor applications.
Instead of designing for one application, we start with common sparse data formats, each of which supports multiple applications.
  Using a \emph{declarative} programming model, \name supports application-independent sparse iteration and memory primitives that can be mapped to vectorized, high-performance hardware.
We optimize random-access sparse memories with configurable out-of-order execution to increase SRAM random-access throughput from 32\% to 80\%.

For a variety of sparse applications, \name with DDR4 memory is \fastercpu\x faster than a multi-core CPU baseline, while \name with HBM2 memory is \fastergpu\x faster than an Nvidia V100 GPU\@.
For sparse applications that can be mapped to Plasticine, a recent dense RDA, \name is 7.6\x to 365\x faster and only \morearea\% larger.
 
\end{abstract}

\maketitle
\section{Introduction}

Sparse linear algebra is used for efficient circuit simulation, finite-element analysis, machine learning (ML), and more~\cite{davis2011university,han2015deep,iandola2016squeezenet}.
Similarly, graph analytics provides insights into large, unstructured datasets like web graphs \cite{leskovec2009community} and social networks \cite{ripeanu2002mapping}.
With data sizes increasing and CPU performance stalled, new architectures are necessary for sparse computation.

With all else equal---process node, die area, and algorithm---a single-purpose ASIC would outperform its programmable counterparts by hyper-specializing its memory hierarchy, data movement, and compute-memory ratio.
However, when building bespoke accelerators, every specialization (e.g., low precision and on-chip memory capacity) is a gamble that ML algorithms will remain unchanged over the roughly five years between design and decommissioning. 
Furthermore, expensive hardware (e.g., HBM, large dice, or leading process nodes) and complicated design steps (e.g., multiple power domains, full-custom datapaths, or low-swing interconnects~\cite{lowswing}) are hard to justify in bespoke accelerators.
Finally, every semi-reconfigurable accelerator introduces a new compiler, which prevents the development of cross-domain applications. 
These factors motivate \emph{unified} accelerators.

Prior accelerators have focused on a few kernels (e.g., BLAS) with hand-written implementations stitched together by end-users. 
However, kernel-driven programming makes it challenging to design new applications and eliminates the opportunity for on-chip kernel fusion~\cite{zhao2019serving, stuckinarut}.
For example, Krylov methods (a building block for optimization, simulation, and scientific computing~\cite{van1992bi}) run multiple sparse and dense kernels which must be fused for efficient execution. 
Kernel-driven programming has also driven ML towards large batches, which have higher inference latency~\cite{jouppi2017datacenter} and worse statistical properties~\cite{keskar2017largebatch}. 

Furthermore, na\"ive fusion of streaming kernels is insufficient because every sparse program is also parameterized by input formats, which exploit problem-specific structure in input data.
   \emph{Compiling} sparse applications avoids writing, optimizing, and testing hundreds of unique kernels; instead, programmers can get optimized code for tensor computations from an algebraic expression and a list of input data formats.
   Recent sparse tensor compilers optimize applications over \emph{sparse iteration spaces,} which are analogous to affine loop nests in dense linear algebra~\cite{kjolstad2020sparse}.
Sparse iteration spaces support any problem expressible as a tensor sum or product, including ML training \emph{and} inference, graph analytics~\cite{kepner2016mathematical}, simulation, and optimization.
Due to the benefits of kernel fusion and input-format specialization, an ideal accelerator for datacenter applications should be \emph{programmable} beyond stitching together individual kernels.

Reconfigurable dataflow architectures (RDAs) are a promising class of accelerators being explored by industry and academia \cite{prabhakar2017plasticine, vissers2019versal, sambanova, cerebras, simplemachines}.
RDAs provide a sea of flexible compute and memory resources in a programmable interconnect, enabling dataflow pipelines that exploit parallelism within and between iterations~\cite{zhao2019serving}. 
Most proposed RDAs have focused on primitives benefiting \emph{dense} workloads, with a vectorized datapath to exploit SIMD parallelism, a simple memory system, and short buffers at each node's input to avoid global pipeline interlocks. However, sparse applications present several challenges to efficient use of SIMD datapaths and on-chip memory bandwidth due to their irregularity and dynamic, data-dependent communication patterns. Ideally, unified \emph{sparse-dense} RDAs should retain efficiency for dense workloads when adding native support for sparse ones.

   Mapping sparsity to dense RDAs is complicated by \emph{structural} hazards (when accessing memory) and \emph{control} hazards (when iterating).
   First, sparse applications frequently use pointer-indexed accesses. 
   Multiple concurrent accesses may point to the same memory bank, even though each bank can only serve one access every cycle.
   Current solutions either tolerate low throughput or over-provision banks and crossbars to minimize conflicts~\cite{dadu2019towards}. 
   Second, sparse iteration may have inter-loop control dependences when iterating over two sparse dimensions.
   Using a scalar programming model~\cite{dadu2019towards}, each non-zero input would have to be compared before deciding whether to dequeue from one list or another, because each comparison's result changes the next comparison's inputs.

   Scheduling accesses to avoid structural hazards is complicated by \emph{positional dataflow,} the typical paradigm for RDAs.
   In positional dataflow, senders and receivers are synchronized, and loop indices are communicated implicitly via the sequence of data elements~\cite{zhang2019scalable}. 
   A compute graph can thus be mapped to parallel, pipelined execution units without reordering data elements or sending control information across the network.
  However, sparsity requires moving a memory request from the originating lane to the correct bank; integration with positional dense computation requires that this shuffling is \emph{precisely} undone.

  In this paper, we introduce \emph{\name,} a positional-dataflow, sparse-dense hybrid RDA that is programmable while approaching bespoke ASICs' performance.
  \name uses a parallel-patterns abstraction for \emph{declarative} sparsity: users express \emph{what} they want to compute, which permits optimized hardware for vectorized sparse iteration and dynamic memory reordering.
  \name supports \emph{all} sparse-iteration tensor applications with a low-level map-reduce programming model based on Spatial~\cite{koeplinger2018spatial} and a clear path to high-level programming (e.g., via TACO~\cite{kjolstad2020sparse}).

\begin{table}
  \centering
  \caption{Some formats for sparse matrices (2-D tensors).}
  \label{fig:tensor_storage}
  \footnotesize
  \begin{tabu}{ll}
    \toprule
    \rowfont{\bfseries\sffamily} Name & Description \\
    \midrule
    CSR & Dense \added{rows}, compressed \added{columns}. \\
    CSC & Dense \added{columns}, compressed \added{rows}. \\
    COO & Compressed non-zeros with row/column pointers. \\
    DCSR & Compressed \added{rows}, compressed \added{columns}. \\
    DCSC & Compressed \added{columns}, compressed \added{rows}. \\
    Banded & Dense along a subset of diagonals. \\
    BCSR & CSR, with $k\times k$ blocks instead of $1\times 1$ non-zeros. \\
    \bottomrule
  \end{tabu}
\end{table}

  \name lessens the impact of memory-bank structural hazards by scheduling on-chip memory accesses over multiple cycles to increase request-level parallelism without increasing crossbar size.
  A flexible memory-shuffle network then increases memory parallelism beyond the number of SIMD lanes while respecting the positional constraints needed for sparse-dense kernel fusion.
  Finally, to enable vectorized sparse iteration in the presence of control hazards, \name uses a scanner to make multiple control-flow decisions per cycle while still supporting a wide range of applications.

\name's key contributions are:
\begin{itemize}
  \item An evaluation of a programmable RDA that supports efficient sparse \emph{and} dense computation, including a comparison against multiple state-of-the-art baselines.
  \item A memory reordering pipeline that exploits the timing flexibility of a loosely-coupled RDA to increase random-access throughput from 32\% to 80\%.
  \item Specialized sparse iteration hardware that exploits declarative sparsity's flexibility to run multiple loop iterations in a single cycle.
\end{itemize}

Our primary baseline design is Plasticine~\cite{prabhakar2017plasticine}, a state-of-the-art dense vector RDA.\@
\name adds only \morearea\% chip area and \morepower\% power, retains its baseline's flexibility, performance, and programmability for dense applications, and outperforms all general-purpose and many application-specific baselines.
\name provides significant performance improvements over Plasticine for sparse applications.
The specific speedup varies depending on the limiting characteristic: structural hazards when reading on-chip memory (17\x), data hazards when modifying memory (SRAM 184\x, DRAM est.\ 1000\x), and control hazards during sparse iteration (est.\ 17\x--131\x faster).
We also compare with an Nvidia V100 GPU and a four-socket Xeon E7-8890v3, using \name with DDR4 memory for the CPU comparison.
\name outperforms the CPU by \cpuminfaster\x to \cpumaxfaster\x and the GPU by \gpuminfaster\x to \gpumaxfaster\x despite using fewer resources.
Finally, \name outperforms state-of-the-art bespoke sparse accelerators (SCNN~\cite{parashar2017scnn}, Graphicionado~\cite{ham2016graphicionado}, and MatRaptor~\cite{srivastava2020matraptor}); \name is slower than EIE~\cite{han2016eie} because it does not have enough SRAM to store matrix data entirely on-chip.

\section{Declarative Tensor Sparsity}
\begin{figure}
  \centering
  \includegraphics[scale=0.70]{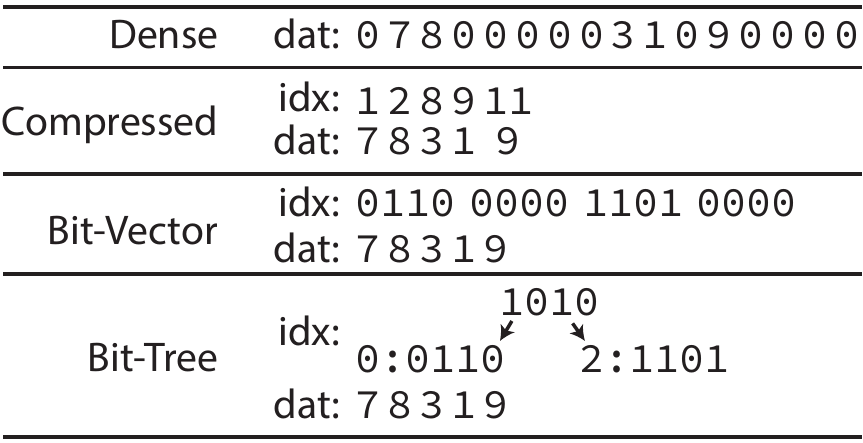}
  \caption{Several formats for sparse vector storage. The bit-vector and bit-tree formats are specialized for implementation using fixed-length memories.}
  \label{fig:vector_storage}
\end{figure}
\begin{table*}
 \centering
 \caption{\name's applications expressed as sparse iteration spaces. Brackets indicate iterations: M[r][c] iterates over rows, then columns, while M[r,c] iterates over non-zeros (rows and columns simultaneously).}
 \label{tab:apps}
 \footnotesize
 \begin{tabu}{clllll>{\scriptsize}l}
 \toprule
 \rowfont{\footnotesize\sffamily\bfseries} App & Data & Format & Loop Over & Iteration & Rand. Accesses & Operation \\
 \midrule
 \multirow{3}{*}{\parbox{.65in}{\centering \sffamily\bfseries CSR SpMV PR-Pull}} & V[c] & Dense & 1. Matrix Rows & dense(r) & & Out[r] = reduced \\
 & M[r][c] & CSR & 2. Cols in Row & dense(len(M[r])) & V[c] & reduced += M[r][c]*V[c] \\
 & Out[r] & Dense & \\ \midrule

 \multirow{3}{*}{\parbox{.65in}{\centering \sffamily\bfseries COO SpMV PR-Edge}} & V[c] & Dense & 1. Matrix Values & dense(nnz(M)) & V[c], Out[r] & Out[r] += M[r,c]*V[c] \\
 & M[r,c] & COO & & \\
 & Out[r] & Dense & \\ \midrule

 \multirow{3}{*}{\sffamily\bfseries CSC SpMV} & V[c] & Dense & 1. Non-Zero Inputs & sparse(V) & M[c] & \\
 & M[c][r] & CSC & 2. Rows in Col & dense(len(M[c])) & Out[r] & Out[r] += M[c][r]*V[c] \\
 & Out[r] & Dense & \\ \midrule

 \multirow{3}{*}{\sffamily\bfseries Conv} & In[iC,r,c] & Dense & 1. Input Non-Zeros & sparse(In) & K[iC] \\
 & K[iC][rK,cK,oC] & Dense/COO & 2. Kernel Non-Zeros & dense(nnz(K[iC])) & Out[...] & Out[oC,r+rK,c+cK] \\
 & Out[oC,r,c] & Dense & & & & ~~+= In[iC,r,c]*K[iC][rK,cK,oC] \\ \midrule

 \multirow{4}{*}{\sffamily\bfseries BFS} & Fr[n] & Bitset & 1. Frontier Nodes & sparse(Fr) & G[s] & \\
 & Rch[n] & Bitset & 2. Adj. Nodes & dense(len(G[s])) & Rch[d], Ptr[d], Fr[d] & Ptr[d] = Rch[d] ? Ptr[d] : s \\
 & G[n][n] & CSC & & & & Fr[d] |= !Rch[d] \\
 & Ptr[n] & Dense & & & & Rch[d] = True \\ \midrule

 \multirow{5}{*}{\sffamily\bfseries SSSP} & Fr[n] & Bitset & 1. Frontier Nodes & sparse(Fr) & G[s], Dist[s] & \\
 & Dist[n] & Dense & 2. Adj. Nodes & dense(len(G[s])) & Dist[d], Ptr[d], Fr[d] & nd = Dist[s] + G[s][d] \\
 & G[n][n] & CSC & & & & Ptr[d] = Dist[d] > nd ? s : Ptr[d] \\
 & Ptr[n] & Dense & & & & Fr[d] |= Dist[d] > nd \\
 & & & & & & Dist[d] = min(Dist[d], nd) \\ \midrule
 \multirow{3}{*}{\sffamily\bfseries M+M} & A[r][c] & CSR-BitTree & 1. Matrix Rows & dense(r) & & C[r].end = reduced + C[r-1].end \\
 & B[r][c] & CSR-BitTree & 2. Cols in Rows ($\bigcup$) & sp-sp(A[r],B[r]) & A[r][c], B[r][c] & reduced += 1 \\
 & C[r][c] & CSR & & & & C[r].push(c, A[r][c] + B[r][c]) \\ \midrule
   \multirow{5}{*}{\parbox{.65in}{\centering \sffamily\bfseries SpMSpM \cite{10.1145/355791.355796}}}
 & A[i][j]  & CSR     & 1. Output Rows   & dense(i)            &          & C[i].end = reduced + C[i-1].end \\
   & B[j][k]  & CSR-Bit & 2. Dim C         & dense(len(A[i])    & B[j] &                  \\
 & C[i][k] & CSR     & 3a. Output Cols & dense(len(B[j]))   & Val[i][k]& Val[i][k] = True \\ 
   & Val[i][k] & Bitset  & 3b. Output Cols ($\bigcap$) & sp-sp(B[j],Val[i]) & C[i][k], B[j][k] & C[i][k] += B[j][k] \\
   &           &         & 3c. Output Cols & sparse(Val[i])      &          & reduced += 1 \\
 \bottomrule
 \end{tabu}
\end{table*}

\name operates on \emph{sparse tensor}s: $k$-dimensional arrays of data with some elements being zero.
The fraction of non-zero elements is the \emph{density,} ranging from around one in two to less than one in one quadrillion.
\name exploits hierarchical parallelism (i.e., along multiple tensor dimensions simultaneously) via loop nests, with multiple iterations of the same loop running in parallel. 

We use a low-level programming model based on \emph{sparse iteration} to express complex computations as linear loops and support a wide variety of tensor formats.
This is a \emph{declarative} model because users do not traverse sparse data structures with comparisons and pointer increments.
Instead, hardware transforms the sparse data-structure into an iterable list of pointers, which permits a higher-performance implementation.

\subsection{Sparse Tensor Formats}
There are many ways to store sparse tensors~\cite{chou2018format}, as shown in \Cref{fig:tensor_storage}.
Formats are specialized based on both the structure of the tensor (are the non-zeros clustered together?) and the operation to be performed (row-first or column-first iteration?).
Any multi-dimensional tensor storage format is a hierarchy of lower-dimensional formats, with the lowest-level formats storing vectors (\Cref{fig:vector_storage}).

For example, consider the venerable compressed-sparse row (CSR) matrix format.
Iterating along rows, the matrix is dense with one entry per row; sparsity is only exploited among columns within a row.
If iteration along rows were sparse, the matrix---with the same row format---would be a doubly-compressed sparse row (DCSR) matrix.
Compressed-sparse column (CSC) also uses one dense and one sparse axis; it permits skipping columns that would be multiplied by zero.
Coordinate (COO) matrices permit iteration only over non-zero tensor values---not rows or columns---with more efficient storage for extremely sparse matrices.
Other formats---especially for vector architectures---use block sparsity (e.g., BCSR), with small (e.g., $16\times16$) dense regions instead of individual elements. 

Finally, some dense vectors (e.g., frontier sets) have boolean elements, motivating a packed bit-vector format.
Bit-vectors can also implicitly point to elements in a compressed array.
\name is designed to support all of these formats; we test it with CSR, CSC, and COO matrices to capture a variety of traversal behaviors.
As described below, we convert compressed vectors to packed bit-vectors (and a bit-tree variant, \Cref{sec:bittree}) for certain operations.

\subsection{Sparse Iteration Spaces}
\label{sec:spiter}
  Map-reduce parallelism breaks every problem into three parts: an iteration space, a pure (no side effects) scalar function, and a reduction function.
  Consider multiplying two dense matrices $A$ and $B,$ with sizes $i\times j$ and $j\times k:$ $C_{ik} = \sum_j A_{ij}B_{jk}.$
  This kernel iterates over three dimensions: $i\times k \times j \Rightarrow A_{ij}B_{jk},$ and the reduction sums along the $j$ dimension to yield a two-dimensional result: $C_{ik}=\sum_jA_{ij}B_{jk}.$
  In a dense system, $i,$ $j,$ and $k$ would be generated by counters, which can be generalized to sparse iteration spaces by making one or more iterated dimensions sparse.
  Kjolstad~\cite{kjolstad2020sparse} provides a comprehensive introduction to sparse iteration spaces; here, we focus on the theory needed to map sparse applications to dataflow architectures.

  Compressed dimensions are the simplest case and can be handled entirely by indirect memory accesses. 
  Assume that $B$ is dense and $A$ has a compressed-sparse row (CSR) format, which is dense along $i$ and sparse along $j.$
  We start by iterating over $i$ and $k,$ which are dense, but only iterate over the subset of $j$ which is non-zero in $A_i.$
  Thus, our $i \times k$ iteration space yields a \emph{compressed} third dimension that uses a counter $j'$ to index $A_i.$ 
  This yields a data value $A_{ij'}$ and a dense (uncompressed) index $j.$
  We are left with dense iteration on a smaller space:
  $i \times k \times\left[j'\rightarrow j\right] \Rightarrow A_{ij'}B_{jk}.$
  
\begin{figure}
  \centering
  \includegraphics{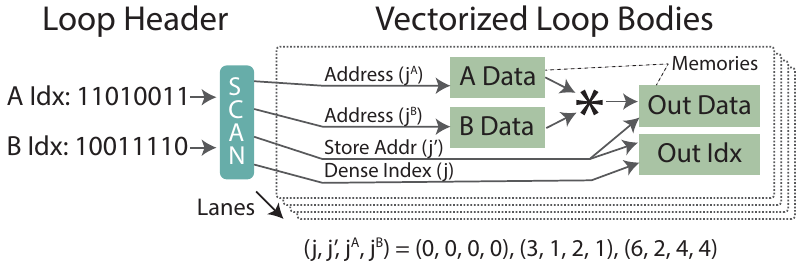}
  \caption{Element-wise vector-vector multiplication using \name. Specialized loop-header hardware divides the problem into parallel loop bodies, which only communicate through atomic accesses and reduction.}
  \label{fig:iterator}
\end{figure}

  Alternately, both $A$ \emph{and} $B$ could be sparse along the $j$ dimension.
  For this case, we introduce a bit-vector format to permit efficient intersection or union computation (\Cref{fig:iterator}).
  Intersecting the $A$ and $B$ bit-vectors yields an iteration space $j',$ which maps to compressed indices $j^A$ and $j^B.$
  The indices $j^A$ and $j^B$ are not guaranteed to equal $j',$ because $j'$ references a dense iteration space (a sequential counter, as in CSR), while $j^A/j^B$ may skip values in the compressed tiles.
  This is the optimal iteration space because it does not multiply by zero: $i\times k\times\left[\mathrm{intersect}(A_i,B_k) \rightarrow j',j^A,j^B\right] \Rightarrow A_{ij^A}B_{j^Bk}.$

Finally, values along the $j$ dimension must be summed.
If the $j$ dimension is the last (innermost) iterated dimension, then all temporary values for one $\left(i,k\right)$ pair will be adjacent and can be summed with a dense reduction.
However, an outer-product GEMM, for example, starts by iterating along $j$ and creates temporary results in random order.
For such cases, \emph{atomic accesses} permit coherent in-place updates: $C_{ik} = C_{ik} + A_{ij}B_{jk}.$

\subsection{Programming \name}
Having identified a map-reduce model for sparse iteration, we need to provide users with an equivalent, easy-to-use programming abstraction.
Spatial~\cite{koeplinger2018spatial} uses a loop-nest abstraction, which lets users nest map-reduce iterations via loop levels.
Our \name dialect extends current loops (\lstinline|Foreach| and \lstinline|Reduce|) by adding a \lstinline|Scan| statement in the place of counters:
\begin{center}
  \small
  \begin{tabular}{r@{\hskip1em}l}
    Dense: &\lstinline|Foreach(min until max by step par p) { j => ...}|\\
    Sparse: &\lstinline|Foreach(Scan(par=p, len=l, A.deq, B.deq))| \\
    &\lstinline|     { j, jA, jB, jprime =>...} |
  \end{tabular}
\end{center}
Users can nest dense loops within sparse, sparse within dense, and so forth.
They can also write arbitrary code (including atomic accesses) in loop bodies and use sparse and dense reductions at multiple levels.
By composing sparse loops natively with dense primitives, \name supports the wide variety of applications shown in \Cref{tab:apps}.
More importantly, by providing an orthogonal, regular programming abstraction, \name is more likely to support future applications.

\paragraph*{Bit-Tree Iteration}
\label{sec:bittree}
Bit-vector sparsity begins to break down when applied to extremely sparse problems (e.g., less than 1\% input density), including our matrix-matrix addition benchmark.
For such problems, \added{sparse iteration} can be nested to support the bit-tree format shown in \Cref{fig:vector_storage}. 
A two-level bit-tree can encode 262,144 zeros with 512 bits.
Streaming iteration on bit-trees is possible using a two-pass algorithm.
In the first pass, sparse-sparse iteration over the top-level vectors \emph{realigns} the lower-level bit-vectors.
In union mode, zeros are inserted to balance unmatched second-level vectors; in intersection mode, unmatched second-level vectors are dropped.
Finally, the top-level vector and second-level vectors are processed by nested sparse-sparse \added{loops}.

For randomly-distributed sparse datasets, bit-tree sparsity would achieve low vectorized throughput.
However, real-world datasets are frequently not randomly distributed: instead, values are clustered near the diagonal or in blocks away from the diagonal.
When values are clustered, bit-tree iteration is able to vectorize across the values in a cluster.

\paragraph*{Memory Ordering Constraints}
\begin{table}
  \footnotesize
  \centering
  \caption{\name's memory ordering modes.}
  \label{tab:ordering}
  \begin{tabu}{ll}
    \toprule
    \rowfont{\bfseries\sffamily} Mode & Access Ordering Contraint \\\midrule
    Unordered & Accesses complete once in arbitrary order. \\
    Address Ordered & Accesses to the same address are ordered.\\
    Fully Ordered & Accesses complete in program order.\\
    \bottomrule
  \end{tabu}
\end{table}

\added{\name offers a choice of memory ordering strictness (\Cref{tab:ordering}).}
For certain cases (e.g., SSSP and deterministic floating-point accumulation), same-address access reordering is \emph{not} permissible.
\added{Programmers can select} address ordering, which permits reordering across addresses but not to the same address.
Finally, certain programs may require that no reordering be performed: full ordering is provided for this edge case, although it is slower than our arbitrated baseline (which reorders accesses within a vectorized request).

\subsection{Case Study: SpMSpM}
Row-based (Gustavson's~\cite{10.1145/355791.355796}) sparse matrix-matrix multiply (SpMSpM) is an asymptotically-efficient algorithm that processes each output row in parallel, which makes it a good candidate for pipelined architectures.
This SpMSpM variant loops along three dimensions, starting with rows in the output, $C,$ and left input matrix, $A.$
Next, it loops over non-zero column entries in $A$'s row $(A_{ij})$ and fetches the corresponding rows from the right matrix $B;$ it multiplies each row $(B_{j})$ by the left matrix's value and (sparsely) adds them to produce the output row:
$C_{ik}=\sum_jA_{ij}B_{jk}.$

When computing each output row on \name, the first step \added{is computing the union of the input rows' bit-vectors}, which yields a bit-vector indicating which entries in $C_i$ will be non-zero.
Then, each input bit-vector is intersected with the output indices; this produces addresses that can be used to accumulate directly into a \emph{compressed} local tile.
Finally, the compressed local tile is swapped with zero (to prepare for the next iteration) and written to DRAM using sparse iteration.
Pre-computing indices may output zeros if two added values sum to zero, but is more efficient and generally accepted~\cite{mkl}.
If zeros in the output are unacceptable, a second sparse iteration pass on $C_i$ can remove them.

\begin{figure*}
  \centering
  \includegraphics[scale=1.000]{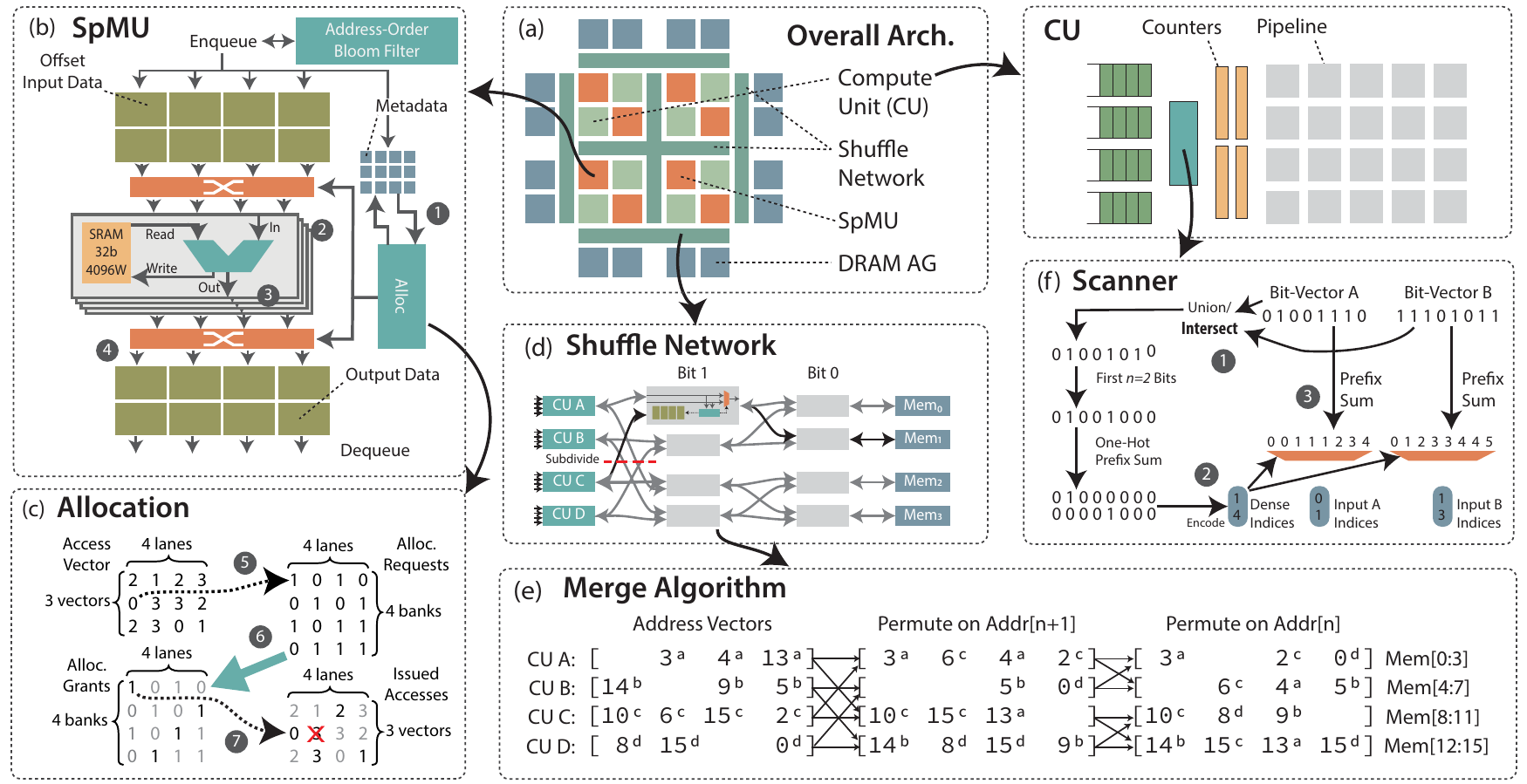}
  \vskip-\baselineskip
  \caption{\name's architecture (a) including: (b) the SpMU pipeline; (c) one iteration of allocation; (d) the shuffle network; (e) vectors of requests being shuffled to the memories; (f) the logical vectorized scan operation.}
  \label{fig:everything}
  \end{figure*}
\section{The \nameupper Architecture}
\label{sec:uarch}
\added{This section describes key architectural features in \name:} dynamically scheduled on-chip sparse accesses, shuffle networks for application outer-parallelization, and scanners for sparse iteration. 
Like Plasticine, \name is built as a checkerboard grid of compute units (CUs) and memory units (MUs) surrounded by address generators (AGs), as shown in \Cref{fig:everything}a.
In this section, we fix \name's lane count $(l=16),$ banks \added{per SpMU} $(b=16),$ and local buffer depth $(d=16).$

\subsection{Sparse Memory Unit (SpMU)}
On-chip sparse accesses are handled by sparse memory units (SpMUs), which dynamically schedule sparse requests to banks (\Cref{fig:everything}b).
\added{The SpMU's main architectural component is a reordering pipeline added to Plasticine's MU.}

\added{Dense programs have a fixed non-conflicting lane-bank mapping, which is typically an identity map: lane 0 to bank 0, and so forth.
However, sparse programs have a random mapping, where multiple lanes may request the same bank; this would require a multi-cycle stall while accesses are resolved.
Therefore, \name introduces a \emph{scheduled} pipeline where $d$ vectors are buffered to stop a single bank conflict from creating a multi-cycle stall.
}

Scheduling \added{the lane-bank crossbar} is non-trivial: up to $l\cdot d$ candidates from $l$ lanes bid \emph{every cycle} for access to $b$ banks.
A greedy solution \added{(lane 0 gets its choice of banks, then lane 1, etc.)} is sub-optimal because one lane-bank matching can block multiple other candidates. 
Making matters worse, head-of-line blocking by straggling requests means that even an exact solution may cause slowdowns later.
To identify the best approximation (one that balances hardware resources and performance), we conduct sensitivity studies with random access traces.

Furthermore, some applications (e.g., Conv) have pathological strided access patterns: with a naive, \emph{linear} bank-mapping scheme, accesses strided by $2^n$ for $n \geq \log_2b$ will hit the same bank and must be serialized.
  Therefore, we \emph{hash} addresses to get a bank ID ($a_{0:3}\oplus a_{4:7} \oplus a_{8:11} \oplus a_{12:15}$) that guarantees that any stride will map to sequential banks.
  Hashing is a common technique for eliminating pathological effects of strided accesses (e.g., in caches~\cite{seznec1993skewed}).

\added{As shown in \Cref{fig:everything}b,} every memory access pending in the issue queue first bids for access to its bank~\circone. \added{Then, an allocator computes a valid crossbar configuration}. 
Following allocation, the SpMU configures its crossbars to route requests from the input queue lanes to banks~\circtwo.
Because the issue queue can only issue one request per lane regardless of queue depth, crossbar size is independent of scheduling depth.
Otherwise, the crossbar needed would grow from $l\times b$ to $l\cdot d \times b.$
The allocator's decisions travel through the pipeline over multiple cycles---the decision made for crossbar traversal in cycle $n$ will control reads in cycle $n+1$ and writes and the output crossbar in cycle $n+2.$ 

Each request then enters an independent read-modify-write (RMW) execution pipeline with one SRAM bank and an FPU (Floating Point Unit)~\circthree, which is capable of integer and floating point addition and subtraction along with several bitwise operations.
The execution unit has separately configurable result muxes for returned data and updated memory values, which allows operations like test-and-set, write-if-memory-zero, swap, min-report-changed, and max.
For example, min-report-changed can be used for SSSP distance updates, and write-if-memory-zero can be used to avoid overwriting backpointers in BFS.
Finally, another crossbar inversely permutes the data based on the bank allocation and writes to the output queue \circfour.
When all requests in a vector complete, the vector is ready to dequeue.

\subsubsection{Allocation}
\added{The detailed allocation process is shown in \Cref{fig:everything}c.
The first stage of allocation is hashing input addresses into bank requests and converting them to one-hot vectors (length $b$).
  The $d$ one-hot vectors in each lane are bitwise-ORed into a vector with \emph{all} requested banks~\circfive; concatenating lanes' requests forms a $l\times b$ request matrix.
}
The request matrix's rows correspond to banks and columns correspond to lanes: if $R_{b,l}=1,$ then there is at least one request from lane $l$ to bank $b.$
A three-iteration, input-first $l\times b$ separable allocator identifies \circsix{} a set of non-conflicting accesses (at most one per lane and one per bank).
If a lane in the issue queue contains multiple requests to the same bank (e.g., lane 1 has two requests for bank 3), a per-lane priority encoder grants the oldest one~\circseven.
The allocator runs in a single cycle: granted requests are immediately marked as issued in a metadata buffer so they do not bid again.

Every \emph{separable allocation}~\cite{becker2009allocator} iteration consists of two stages of fixed-priority arbiters.
The first stage prunes the matrix so that every lane requests at most one bank, and the second stage ensures that every bank selects at most one lane.
These two pruning steps guarantee at most one grant per bank and lane.
However, if the first iteration chooses suboptimally, more grants could be added. 
Successive stages consider requests that were not previously granted \emph{and} do not conflict with established grants.

Our allocator prioritizes older requests because they can cause head-of-line blocking.
In \name's 16-{slot queue}, the first {five} slots bid in the first round of allocation, the first ten {slots} bid in the second round, and all bid in the third.
The performance benefits of multiple priorities are shown in \Cref{tab:spmu_prio}, with every design point using three allocation iterations; for configurations with fewer than three priorities, the final iterations use all requests.

\added{The degree of the SpMU's reordering is visible in \Cref{fig:order}.
Due to four requests for bank 10, the arbitrated baseline executes this vector over four cycles.
The unordered SpMU has \emph{higher latency} for a single vector, but \emph{higher throughput} as well: the pale-gray numbers show reordered requests from before and after the traced vector.
}

\begin{table}
  \footnotesize
  \centering
  \caption{SpMU throughput (the percentage of banks active per cycle) increases by using deeper queues and more priorities.}
  \label{tab:spmu_prio}
  \begin{tabu}{rrrrrr}
    \toprule
    \rowfont{\bfseries\sffamily}       &         & & \multicolumn{3}{c}{Bank Use (\%)} \\ \cmidrule{4-6}
    \rowfont{\bfseries\sffamily} Depth & \added{Crossbar} & Sched. (\si{\micro m^2}) & 1-Pri.                             & 2-Pri.  & 3-Pri. \\ \midrule
                            8          & \added{16\texttimes16} & \num{38052}            & 51.5                            & 66.4 & 67.9  \\
                            8          & \added{32\texttimes16} & \num{48938}            & 55.3                            & 68.5 & 72.5  \\ \midrule
                           16          & \added{16\texttimes16} & \bfseries \num{51359}            & 63.9                            & 79.9 & \bfseries 79.9  \\
                           16          & \added{32\texttimes16} & \num{62918}            & 67.8                            & 85.1 & 85.4  \\ \midrule
                           32          &  \added{16\texttimes16}& \num{79301}            & 72.7                            & 84.7 & 84.7  \\
                           32          &  \added{32\texttimes16}& \num{90433}            & 77.0                            & 92.4 & 92.5  \\
    \bottomrule
  \end{tabu}
\end{table}

\subsubsection{SpMU Design Space}
\begin{figure}[tb]
  \resizebox{\linewidth}{!}{%
  \scriptsize%
  \centering%
  \setlength{\tabcolsep}{2pt}%
  \newcommand{\prioaddr}[1]{\textcolor{solarized-base02}{\textbf{#1}}}
  \newcommand{\otheraddr}[1]{\textcolor{solarized-base0}{#1}}%
  \newcommand{\labbox}[2]{\parbox{.75in}{\centering {\sffamily\bfseries #1}\\Util: #2}}%
  \begin{tabular}{l@{\hskip2pt}c@{\hskip2pt}rrrrrrrrrrrrrrrrrrrr}
    \toprule
    & \footnotesize\sffamily Cyc. & \multicolumn{16}{c}{\footnotesize\sffamily Lanes (0--15)}\\
    \midrule
    \multirow{15}{*}{\labbox{Unordered}{79.9\%}} &0 &\prioaddr{4} &\otheraddr{14} &\otheraddr{8} &\otheraddr{11} &\prioaddr{3} &\otheraddr{9} &\otheraddr{0} & &\otheraddr{7} & &\otheraddr{15} &\otheraddr{13} & &\otheraddr{10} &\otheraddr{6} &\otheraddr{12}\\
    &1&\otheraddr{1} &\otheraddr{14} &\otheraddr{4} &\otheraddr{12} &\otheraddr{8} &\otheraddr{13} &\otheraddr{2} &\otheraddr{9} & &\otheraddr{0} & &\otheraddr{11} &\otheraddr{6} &\otheraddr{10} &\otheraddr{7} &\otheraddr{3}\\
    &2&\otheraddr{5} &\prioaddr{1} &\otheraddr{14} &\otheraddr{10} & &\otheraddr{12} &\otheraddr{8} &\otheraddr{13} &\otheraddr{9} &\otheraddr{11} & &\otheraddr{4} &\otheraddr{2} &\otheraddr{7} &\otheraddr{15} &\otheraddr{3}\\
    &3&\otheraddr{1} &\otheraddr{3} &\otheraddr{0} &\otheraddr{5} &\otheraddr{10} &\otheraddr{13} &\prioaddr{7} & & &\otheraddr{14} &\otheraddr{4} &\otheraddr{6} &\otheraddr{11} &\otheraddr{12} &\otheraddr{9} &\otheraddr{8}\\
    &4&\otheraddr{11} &\otheraddr{6} &\otheraddr{9} &\otheraddr{13} & & &\otheraddr{8} &\otheraddr{12} &\otheraddr{10} &\otheraddr{4} & &\otheraddr{14} &\otheraddr{7} &\otheraddr{15} & &\prioaddr{5}\\
    &5&\otheraddr{5} &\otheraddr{11} &\otheraddr{4} & &\otheraddr{9} & &\otheraddr{6} &\otheraddr{1} &\otheraddr{8} &\otheraddr{7} &\otheraddr{12} & & &\otheraddr{0} &\otheraddr{14} &\otheraddr{10}\\
    &6&\otheraddr{14} & &\otheraddr{7} & &\otheraddr{11} & & &\otheraddr{5} &\otheraddr{6} &\otheraddr{10} &\otheraddr{12} &\otheraddr{1} &\otheraddr{9} &\otheraddr{2} &\otheraddr{8} &\\
    &7&&\otheraddr{14} &\otheraddr{1} & &\otheraddr{11} &\otheraddr{9} &\otheraddr{7} &\otheraddr{6} &\prioaddr{2} &\otheraddr{15} &\prioaddr{10} & & &\otheraddr{5} &\otheraddr{8} &\otheraddr{12}\\
    &8&&\otheraddr{11} &\otheraddr{14} & & & &\otheraddr{13} & & &\prioaddr{12} & & &\otheraddr{8} &\otheraddr{4} &\otheraddr{10} &\\
    &9&& & &\prioaddr{12} &\otheraddr{14} &\prioaddr{2} & &\prioaddr{11} & &\otheraddr{0} &\otheraddr{6} & &\prioaddr{10} &\prioaddr{9} &\otheraddr{13} &\otheraddr{8}\\
    &10&& &\prioaddr{8} &\otheraddr{4} &\otheraddr{15} &\otheraddr{5} &\otheraddr{0} &\otheraddr{3} & & &\otheraddr{11} &\otheraddr{12} &\otheraddr{7} & &\prioaddr{10} &\otheraddr{14}\\
    &11&\otheraddr{2} &\otheraddr{7} & &\otheraddr{8} & &\otheraddr{11} &\otheraddr{14} &\otheraddr{15} &\otheraddr{10} &\otheraddr{3} & & & &\otheraddr{12} & &\\
    &12&\otheraddr{7} &\otheraddr{12} & &\otheraddr{11} & &\otheraddr{13} &\otheraddr{8} &\otheraddr{5} & &\otheraddr{0} &\otheraddr{10} &\otheraddr{14} &\otheraddr{6} & &\otheraddr{2} &\\
    &13&\otheraddr{13} & &\otheraddr{9} &\otheraddr{12} &\otheraddr{11} &\otheraddr{4} &\otheraddr{10} &\otheraddr{15} & &\otheraddr{3} &\otheraddr{14} &\otheraddr{8} &\otheraddr{7} &\otheraddr{5} & &\otheraddr{2}\\
    &14&\otheraddr{8} &\otheraddr{5} &\otheraddr{7} &\otheraddr{12} &\otheraddr{3} &\otheraddr{2} & &\otheraddr{0} &\otheraddr{11} & &\otheraddr{6} &\prioaddr{10} &\otheraddr{4} &\otheraddr{13} & &\\

    \midrule
    \multirow{6}{*}{\labbox{Address\\Ordered}{34.2\%}}&0& \prioaddr{4} &\prioaddr{1} &\prioaddr{8} &\prioaddr{12} &\prioaddr{3} &\prioaddr{2} &\prioaddr{7} &\prioaddr{11} & & & & & & & &\\
    &1&& & & & & & & &\prioaddr{2} &\prioaddr{12} &\prioaddr{10} & & &\prioaddr{9} & &\prioaddr{5}\\
    &2&\otheraddr{5} &\otheraddr{11} &\otheraddr{7} &\otheraddr{8} &\otheraddr{15} &\otheraddr{9} & &\otheraddr{1} & &\otheraddr{0} &\otheraddr{6} &\otheraddr{12} &\prioaddr{10} & & &\otheraddr{3}\\
    &3&& & & & & &\otheraddr{8} & & & & &\prioaddr{10} &\otheraddr{7} &\otheraddr{12} &\otheraddr{15} &\\
    &4&\otheraddr{1} &\otheraddr{6} &\otheraddr{9} &\otheraddr{11} &\otheraddr{10} & & &\otheraddr{5} &\otheraddr{8} &\otheraddr{3} & & &\otheraddr{2} &\otheraddr{12} & &\\
    &5&& & & & &\otheraddr{9} &\otheraddr{6} & & & &\otheraddr{11} &\otheraddr{1} & & &\prioaddr{10} &\otheraddr{8}\\
    \midrule
    \multirow{5}{*}{\labbox{Fully\\Ordered}{25.5\%}} &0&\prioaddr{4} &\prioaddr{1} &\prioaddr{8} &\prioaddr{12} &\prioaddr{3} &\prioaddr{2} &\prioaddr{7} &\prioaddr{11} & & & & & & & &\\
    &1&& & & & & & & &\prioaddr{2} &\prioaddr{12} &\prioaddr{10} & & & & &\\
    &2&& & & & & & & & & & &\prioaddr{10} & & & &\\
    &3&& & & & & & & & & & & &\prioaddr{10} &\prioaddr{9} & &\\
    &4&& & & & & & & & & & & & & &\prioaddr{10} &\prioaddr{5}\\
  \midrule
    \multirow{4}{*}{\labbox{Arbitrated}{32.4\%}} &0&\prioaddr{4} &\prioaddr{1} &\prioaddr{8} &\prioaddr{12} &\prioaddr{3} &\prioaddr{2} &\prioaddr{7} &\prioaddr{11} & & &\prioaddr{10} & & &\prioaddr{9} & &\prioaddr{5}\\
    &1&& & & & & & & &\prioaddr{2} &\prioaddr{12} & &\prioaddr{10} & & & &\\
    &2&& & & & & & & & & & & &\prioaddr{10} & & &\\
    &3&& & & & & & & & & & & & & &\prioaddr{10} &\\
    \bottomrule
  \end{tabular}
  }
  \caption{ \added{A traced request vector (bold) in a stream of random requests (light font), showing the bank granted for each lane in every cycle.
  Unordered execution executes requests from the vector over more cycles but provides the best throughput with interleaving.
  }
  }
  \label{fig:order}
\end{figure}

The SpMU has several additional optimizations and tunable parameters.
\paragraph*{\added{Crossbar Size}}
In most banked memory systems, arbitration is simplified by overprovisioning banks.
However, with allocation, adding crossbar inputs can also increase throughput (\Cref{tab:spmu_prio}).
Banking the input queue to feed a $2\cdot l \times b$ crossbar (\added{i.e.,} 2\x input speedup~\added{\cite{dally2004principles}}) improves performance slightly but adds \SI{11559}{\micro m^2}.
For our final design, we choose a 16-entry queue without speedup to balance area and performance; however, it is interesting that the 16-entry queue with speedup is faster than the 32-entry queue while consuming less area.

\paragraph*{Address Ordering}
\added{The SpMU's pipeline performs arbitrary reordering of requests, which is typically (but not always) permissible.
To ensure address-based ordering, the SpMU must stall requests \emph{before} they enter the reordering pipeline if they may conflict with in-flight requests.
}
\added{The first step is splitting request vectors if two lanes request the same address.}
Splitting is visible in \Cref{fig:order} as the first access to bank 10 being stalled by one cycle due to a split at bank 2. 
\added{Next, a 128-entry Bloom filter checks for potential conflicts with pending in-queue requests.}
Using 128 entries provides reasonable performance for this less-common access mode while minimally increasing area.

\paragraph*{Repeated-Read Elision}
During enqueue, duplicate read-only accesses are squashed and replaced with metadata indicating the lane of the initial (un-squashed) access. 
When the vector is dequeued, duplicate lanes fill from the one read that performed.

\subsection{Shuffle Network}
\label{sec:uarchshuffle}
Shuffle networks (\Cref{fig:everything}d) combine requests between parallel outer-loop iterations while respecting structural hazards and ordering constraints. 
Each is built out of merge units arranged in a butterfly topology. 
\name has three vertical networks and three horizontal networks: the outer networks connect accesses to DRAM, and the inner networks are used for SRAM accesses.
By disabling links crossing the center of the butterfly network (the red dashed line in \Cref{fig:everything}d), each shuffle network can be divided into two smaller ones (\added{with further recursive division possible as well}).
A bypass path (not shown) detects requests that would not be shuffled (e.g., those from CU 0 to memory 0) and forwards them directly; this lowers load on the shuffle network and makes the lane-wise merging more likely to succeed.

Then, each merge unit takes two vectors of incoming requests and tests a single address bit that determines whether they are forwarded to its half or dropped (i.e., intended for the other half).
Then, the merge unit combines the vectors, shuffling valid entries by up to one lane in either direction.
\added{\Cref{fig:everything}e shows an example shuffle, moving from left to right.
Each CU generates its own vector of addresses, which have superscripts showing the originating CU's letter (A--D) so they can be tracked through the process.
The first merge step does a top-half/bottom-half split, partitioning on Addr[3], and the second merge step partitions on Addr[2]. 
Each stage tracks the shuffle decisions used to merge vectors and inverts these permutations when replies are returned.
}

In addition to keeping the logic small, the restricted shuffling ($\pm$ one lane) decreases the amount of storage for inverse permutation: the merge unit tracks its decisions in a 48-bit (3 bits per lane), 64-entry FIFO.
Keeping a deep buffer is important for tolerating long memory access latencies (DRAM or scheduled SRAM).
On-chip streaming pipelines \emph{flow through} the shuffle network and corresponding memories, so each cycle additional of memory latency can be tolerated by adding another inverse-permutation FIFO slot.

\subsection{Sparse Loop Headers: Scanner}
\begin{table}
  \footnotesize
  \caption{Area of various scanner configurations ($\si{\micro m^2}$).}
  \label{tab:scanarea}
\sisetup{                           
        round-mode      = places,   
        round-precision = 0,        
        per-mode        = symbol,   
      group-four-digits = true,     
      detect-weight = true,
      detect-inline-weight = math,
table-format=6.0,
        }
  \centering
  \begin{filecontents*}{csv/area_reformat.csv}
Width,1,2,4,8,16
1,12.0,13.0,13.0,14.0,17.0
2,25.0,28.0,33.0,45.0,67.0
4,56.0,72.0,100.0,156.0,268.0
8,111.0,147.0,204.0,313.0,531.0
16,597.0,724.0,949.0,1390.0,2369.0
32,834.0,1030.0,1364.0,2112.0,3508.0
64,1272.0,1581.0,2077.0,3150.0,5256.0
128,2157.0,2765.0,3645.0,5591.0,9456.0
256,3985.0,5231.0,6927.0,10674.0,19898.0
512,7777.0,10447.0,14377.0,22562.0,42997.0
\end{filecontents*}
\pgfplotstabletypeset[
    col sep=comma,
    string type,
    column type={S},
    every row 8 column 5/.style={normpoint}, 
    assign column name/.style={
      /pgfplots/table/column name={\multicolumn{1}{r}{\sffamily\textbf{#1}}}
    },
    every head row/.style={before row={\toprule & \multicolumn{5}{c}{\sffamily\textbf{Output Vectorization}}\\\cmidrule{2-6}}, after row=\midrule},
    every last row/.style={after row=\bottomrule},
    skip rows between index={0}{4},
    skip rows between index={4}{7},
    ]{csv/area_reformat.csv}
\end{table}

The scanner, \added{which implements sparse loop headers,}  is a relatively simple block: the key insight is that it requires $O(\log n)$ levels of logic, which is less than the $O(n)$ levels that would be required to run arbitrary independent decisions (e.g., stream join~\cite{dadu2019towards}).
The simplest scanner is the data scanner, which identifies one 32-bit non-zero element in a 16-element vector per cycle.
Because the data scanner can only scan 16 elements per cycle, vectorization could not out-perform dense computation; therefore, the data scanner is not used in inner loops.

The bit-vector scanner (\Cref{fig:everything}f), which is used for vectorized sparse iteration, starts by computing either the intersection or union of its inputs~\circone.
It then identifies the first 16 set bits, which, in a single cycle, it passes into a pipeline and then clears.
Because this is a fairly simple operation, the scanner can easily process one input per cycle.

In the pipeline, the 16 selected bits are passed into encoders~\circtwo{} as a 256\x{}16 array to produce the \added{dense indices ($j$)}.
Finally, the scanner uses these indices to index prefix sums over the inputs and provide indices into compressed input vectors \added{($j^A/j^B$)}~\circthree; if the bit was not set in the input (union mode only), $-1$ is returned as the index.
\Cref{tab:scanarea} shows how scanner area varies with width and vector length.
By choosing a 256-input, 16-output scanner, we use 54\% less area than the largest option, 512\x{}16, which also fails to meet timing.

Scanning is performed before values enter the counter chain: in the first cycle, the scanner outputs the number of valid elements.
This count is written into the appropriate counter and used for iteration, which avoids the need to replicate one scanner per counter level.
For programs that nest more than one scanner, a CU can be used in a scanner-only mode to feed a second CU.
This scanner-only CU also reduces vector input buffer requirements: the follower CU only needs one vector buffer to hold packed indices, instead of two buffers to hold bit vectors.
Finally, SpMUs do not have the full scanner provisioned, but instead have a single-input, scalar-output scanner that adds less area.
When a complicated scan operation is needed for address generation, a CU must be used instead.

\subsection{Off-Chip Access Support}

\paragraph*{Atomic DRAM Accesses}
\name's atomic DRAM support uses a similar pipeline to the on-chip SRAM and is present in every DRAM address generator (AG).
The AG tracks the current status of outstanding bursts; when a new request vector arrives, each access is checked against pending bursts and issued if necessary.
After executing the relevant accesses, the burst is written back to DRAM, ensuring that no reads race writes---if a read would race a write, it is instead marked as pending and executed when the write returns.
To parallelize DRAM accesses, the shuffle network ensures that each AG is responsible for a mutually-exclusive memory region.
Random-access DRAM bandwidth is limited by the memory controller and technology, so \name's DRAM AGs are less sensitive to architectural parameters.

\paragraph*{Compressed Dense DRAM}
Read-only DRAM compression is used to optimize applications that read tiles of pointers (which frequently have closely-spaced values).
\name uses a packet-based memory compression format, with each burst encoded using a base/offset format; a one-byte header specifies the base and offset sizes.
Unlike GPUs, which compress framebuffers in large tiles without programmer input, \name requires pre-compression and restricts compressed loads to tile boundaries.
Removing support for writes and random reads permits denser compression and saves area.

\paragraph*{Format Conversion}
Finally, format-conversion hardware generates bit-vector formats from pointers.
\name's iterators use bit-vector sparsity for computing intersections. 
However, these can be less bandwidth-efficient than compressed pointers.
Converting compressed pointers to bit-vectors in the SpMU would require multiple modifications to the same word, causing bank conflicts and slowing execution. 
Therefore, special-purpose format conversion hardware is added to the compute tile with minimal area overhead.

\begin{table}
  \newcommand{\sep}{$\bullet$}
  \footnotesize
  \centering
  \caption{The datasets used to evaluate \name. Convolution dimensions are dim\sep kdim\sep inCh\sep outCh and activations\sep kernel for non-zeros and density.}
  \label{tab:lindata}
  \begin{tabu}{crrrrrr}
    \toprule
     \rowfont{\sffamily\bfseries}                              & Name                                      & Dim.    & Non-Zeros     & \% Dense\\ \midrule
                                  \multirow{3}{*}{\sffamily\bfseries \rotatebox[origin=c]{90}{SpMV}}
                                  \multirow{3}{*}{\sffamily\bfseries \rotatebox[origin=c]{90}{M+M}}
                                  \multirow{3}{*}{\sffamily\bfseries \rotatebox[origin=c]{90}{\tiny BiCGStab}}
                                                                                            & ckt11752\_dc\_1 \cite{davis2011university}                          & \num{49702}           & \num{333029}    & \num{0.014} \\
                                                                                            & Trefethen\_20000                         \cite{davis2011university} & \num{20000}           & \num{554466}    & \num{0.139} \\ 
                                                                                            & bcsstk30 \cite{davis2011university}                                 & \num{28924}           & \num{2043492}   & \num{0.244} \\
                                                                                            \midrule
                                  \multirow{3}{*}{\sffamily\bfseries \rotatebox[origin=c]{90}{PR}}
                                  \multirow{3}{*}{\sffamily\bfseries \rotatebox[origin=c]{90}{BFS}}
                                  \multirow{3}{*}{\sffamily\bfseries \rotatebox[origin=c]{90}{SSSP}}
                                                                                            & usroads-48 \cite{davis2011university}                               & \num{126146}          & \num{323900}    & \num{0.002} \\
                                                                                            & web-Stanford \cite{leskovec2009community}                           & \num{281903}          & \num{2312497}   & \num{0.003} \\
                                                                                            & flickr \cite{davis2011university}                                   & \num{820878}          & \num{9837214}   & \num{0.001} \\ \midrule
                                  \multirow{3}{*}{\sffamily\bfseries \rotatebox[origin=c]{90}{\tiny SpMSpM}}
                                                                                            & spaceStation\_4            \cite{davis2011university}               & \num{950}             & \num{14158}     & \num{1.6}   \\
                                                                                            & qc324\cite{davis2011university}                                     & \num{324}             & \num{27054}     & \num{25.7}  \\
                                                                                            & mbeacxc       \cite{davis2011university}                            & \num{496}             & \num{49920}     & \num{20.3}  \\
                                  \midrule
                                  \multirow{3}{*}{\sffamily\bfseries \rotatebox[origin=c]{90}{Conv}}
                                                                                            & ResNet-50 ~\#1~\cite{he2016deep}                                    & 56\sep1\sep64\sep64   & 88837\sep1229   & \num{44.3}\sep\num{30.0}  \\
                                                                                            & ResNet-50 ~\#2~\cite{he2016deep}                                    & 56\sep3\sep64\sep64   & 47574\sep11057  & \num{23.7}\sep\num{30.0}  \\
                                                                                            & ResNet-50 \#29~\cite{he2016deep}                                    & 14\sep3\sep256\sep256 & 41552\sep176460 & \num{82.8}\sep\num{30.0}  \\
    \bottomrule
  \end{tabu}
\end{table}

\section{Evaluation}
\label{sec:eval}

We evaluate \name's performance using a custom cycle-accurate C++ simulator, which has previously been used to evaluate other Plasticine-based RDAs~\cite{zhang2019scalable,vilim2020gorgon,zhang2021sara}.
Our simulator models the effects of a hybrid static-dynamic network~\cite{zhang2019scalable} and uses Ramulator~\cite{kim2015ramulator} to model DRAM behavior. 
Applications are written in a dialect of the Spatial language and simulated using the datasets shown in \Cref{tab:lindata}; for DRAM bandwidth and scanner sensitivity studies, {p2p-Gnutella31}~\cite{leskovec2009community} is substituted for {flickr} to make simulation more feasible.
Datasets with many nodes (e.g., flickr~\cite{leskovec2009community}) stress the performance of cross-tile applications---important when scaling out to larger graphs.
To evaluate convolution, we train a ResNet-50 model and prune it to be 30\% dense.
For CSC SpMV (sparse matrix-vector multiplication), we use a 30\%-dense input vector (based on the datasets used to test EIE~\cite{han2016eie}).
Graph datasets are tiled using Metis~\cite{karypis1998fast} with nodes weighted by edge count to give load-balanced tiles.
Linear algebra datasets are tiled using a round-robin division of rows, columns, or non-zero matrix values.

\paragraph*{Convolution Mapping}
All of our evaluated applications can be mapped using Spatial.
However, due to halo exchange, convolution maps poorly to positional dataflow: in a streaming-positional architecture, each tile's accumulation buffer would need eight links (input and output) to neighboring tiles.
Although we can map convolution to the shuffle network using Spatial (using 100\% of the on-chip shuffle resources), using the dynamic network~\cite{zhang2019scalable} in a non-positional (i.e., out-of-order) mode yields 3.8\x higher performance.
Without manual mapping, \name still outperforms a CPU and GPU; however, manual mapping is used to compare against SCNN (which uses a similar tiled architecture).
\begin{figure*}
  \centering
  \pgfplotstableread[col sep=comma]{csv/bw.csv}\tabbw
  \pgfplotstableread[col sep=comma]{csv/area_eff.csvT}\tabarea
\pgfplotstableread[col sep=comma]{csv/bw_comp.csv}\tabcomp
  \begin{tikzpicture}
    \begin{groupplot}[
      groupbase,
    height=1.40in,
  axis x line*=bottom,
  axis y line*=left,
]
        \nextgroupplot[ title = {\added{(a)} DRAM Bandwidth Sensitivity},
  ymode=log,
  xmode=log,
    legend style={
      legend columns=5,
    },
    set layers,
    font=\footnotesize\sffamily,
    xmin=20,
    xmax=2000,
    xtick={10,20,50,100,200,500,1000,2000},
    xticklabels={10,20,50,100,200,500,1000,2000},
    ymin=1.0,
    ymax=100,
    ytick={1,10,100},
    mark options={draw opacity=0, scale=0.5},
  ylabel={Speedup},
  xlabel={Bandwidth (\si{GB/s})},
            legend style = { column sep = 2pt, legend columns = 11, legend to name = grouplegend,}, legend style={draw=none, mark options={draw opacity=0}, font={\footnotesize}}]
            \addplot table[x=key, y=CSR]  {\tabbw};\addlegendentry{CSR}
            \addplot table[x=key, y=COO]  {\tabbw};\addlegendentry{COO}
            \addplot table[x=key, y=CSC]  {\tabbw};\addlegendentry{CSC}

            \addplot table[x=key, y=Conv]  {\tabbw};\addlegendentry{Conv}
            \addplot table[x=key, y=PRPull]  {\tabbw};\addlegendentry{PRPull}
            \addplot table[x=key, y=PREdge]  {\tabbw};\addlegendentry{PREdge}

            \addplot table[x=key, y=BFS]  {\tabbw};\addlegendentry{BFS}
            \addplot table[x=key, y=SSSP]  {\tabbw};\addlegendentry{SSSP}
            \addplot table[x=key, y=MMAddTree]  {\tabbw};\addlegendentry{M+M}
            \addplot table[x=key, y=RowGEMM]  {\tabbw};\addlegendentry{SpMSpM}
        \nextgroupplot[title = {\added{(b)} Area Sensitivity},
  ymode=log,
  xmode=log,
    legend style={
      legend columns=5,
    },
    set layers,
    font=\footnotesize\sffamily,
    xmin=1,
    xmax=100,
    xtick={1,10,100},
    xticklabels={1,10,100},
    ymin=1,
    ymax=12,
    ytick={1,10,100},
    mark options={draw opacity=0, scale=0.5},
  ylabel={Speedup},
  xlabel={Weighted On-Chip Area (\%)},
]

  \addplot table [y=CSR_speedup,       x expr=100*\thisrow{CSR_size}]       {\tabarea};
  \addplot table [y=COO_speedup,       x expr=100*\thisrow{COO_size}]       {\tabarea};
  \addplot table [y=CSC_speedup,       x expr=100*\thisrow{CSC_size}]       {\tabarea};
  \addplot table [y=Conv_speedup,      x expr=100*\thisrow{Conv_size}]      {\tabarea};
  \addplot table [y=PRPull_speedup,    x expr=100*\thisrow{PRPull_size}]    {\tabarea};
  \addplot table [y=PREdge_speedup,    x expr=100*\thisrow{PREdge_size}]    {\tabarea};
  \addplot table [y=BFS_speedup,       x expr=100*\thisrow{BFS_size}]       {\tabarea};
  \addplot table [y=SSSP_speedup,      x expr=100*\thisrow{SSSP_size}]      {\tabarea};
  \addplot table [y=MMAddTree_speedup, x expr=100*\thisrow{MMAddTree_size}] {\tabarea};
        \nextgroupplot[title = {\added{(c)} Compression Sensitivity},
    xmode=log,
    set layers,
    font=\footnotesize\sffamily,
    xmin=20,
    xmax=2000,
    xtick={10,20,50,100,200,500,1000,2000},
    xticklabels={10,20,50,100,200,500,1000,2000},
    ymin=1.0,
    ymax=1.7,
    mark options={draw opacity=0, scale=0.5},
  ylabel={Speedup},
  xlabel={Bandwidth (\si{GB/s})},
]
  \addplot table[x=key, y=CSR]  {\tabcomp};
  \addplot table[x=key, y=COO]  {\tabcomp};
  \addplot table[x=key, y=CSC]  {\tabcomp};
  \pgfplotsset{cycle list shift=1}
  \addplot table[x=key, y=PRPull]  {\tabcomp};
  \addplot table[x=key, y=PREdge]  {\tabcomp};

  \addplot table[x=key, y=BFS]  {\tabcomp};
  \addplot table[x=key, y=SSSP]  {\tabcomp};
    \end{groupplot}
    \node at ($(group c2r1) + (0,2.0cm)$) {\ref{grouplegend}}; 

  \end{tikzpicture}
  \caption{The impact of system-level parameters on \name's performance.}
  \label{fig:system}
\end{figure*}
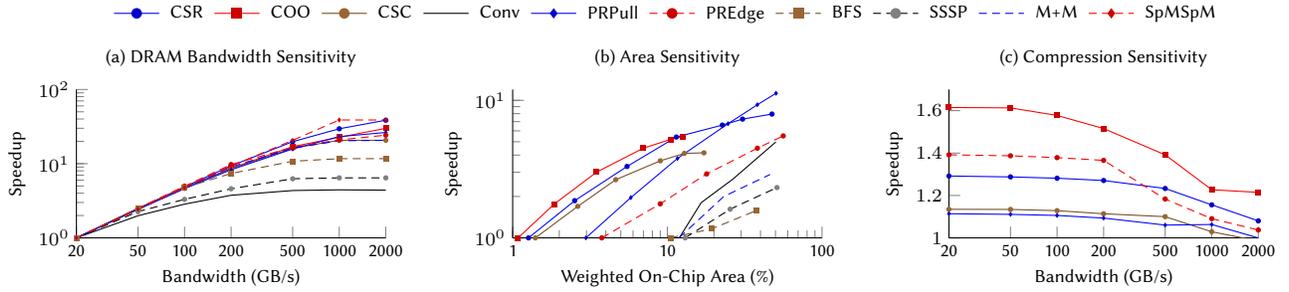

\begin{table}
  \footnotesize
  \centering
  \caption{\name's design parameters.}
  \label{tab:design}
  \begin{tabular}{llr}
    \toprule
    \multirow{3}{*}{\sffamily\bfseries Off-Chip Bandwidth} & HBM-2E & \SI{1800}{GB/s}\\
    & HBM-2 & \SI{900}{GB/s}\\
    & DDR4-2133 & \SI{68}{GB/s}\\
    \midrule
    \multirow{3}{*}{\sffamily\bfseries Grid Size}&Compute Unit& 200 \\
    & Sparse Memories (SpMU)& 200 \\
    & Address Generators & 80 \\
    \midrule
    \multirow{2}{*}{\sffamily\bfseries SpMU}& Banks & 16 \\
    &Capacity& \SI{256}{KiB} \\
    \midrule
    \multirow{2}{*}{\sffamily\bfseries Shuffle Network}& On-Chip& 2$\times$16 \\
    & Off-Chip& 4$\times$16 \\
    \bottomrule
  \end{tabular}
\end{table}

\subsection{\name Architectural Parameters}
We use Plasticine, a recent dense RDA~\cite{prabhakar2017plasticine}, to evaluate \name's hardware costs.
Both \name and Plasticine are programmed using the Spatial language~\cite{koeplinger2018spatial} and SARA low-level compiler~\cite{zhang2021sara}.
Because modifying Plasticine could lower performance for dense applications, we start with binding assumptions about several parameters: vector length, grid layout, and on-chip SRAM banks and dimensions (\Cref{tab:design}).

\paragraph*{Gridded Architecture}
Our first assumption is a 1:1 ratio of homogeneous compute (CU) and memory units (MU).
These form a $20\times20$ checkerboard array, ringed by 80 DRAM address generators (AG). 
AGs send burst-level (\SI{64}{B}) requests to a global controller, which performs low-level DRAM scheduling (precharge, row open, etc.).
Units are connected by a loosely-timed interconnection network with per-link buffering to avoid global synchronicity; it provides vector (512-bit) and scalar (32-bit) links for efficient mapping~\cite{zhang2019scalable}.
Network buffering provides timing flexibility for \name's reordered memory accesses, which may be delayed for several cycles to increase available parallelism.

\paragraph*{Flexible Parallelism}
Each CU has 16 vector lanes and 6 vector stages; stages perform a map or a reduce operation on 32-bit fixed- or floating-point data.
Loops can be parallelized at two levels: within a vector (inner-par) and across multiple vectorized CUs (outer-par).
Loops execute at most once per cycle, so an iteration count not divisible by 16 will leave inactive lanes.
Finally, streaming inter-CU pipelines provide another form of parallelism.

\paragraph*{On-Chip Memory}
\label{sec:plastmem}
On-chip memories are arranged as 16 banks of 4096 32-bit words each, with \SI{256}{KiB} per memory (\SI{50}{MiB} total); each bank has one read and one write port.
Memories are locally addressed, permitting parallelization without a global crossbar.
As described in \Cref{sec:uarch}, each memory is scheduled using a three-iteration allocator.

\subsection{Hardware Resources}
\begin{table}
  \footnotesize
  \centering
  \caption{\name's area relative to Plasticine.}
  \label{tab:area}
  \newcommand{\tabin}{\hspace{1em}}
  \newcommand{\sepelem}{\\[-0.85em]}
  \begin{tabu}{lrrrrrr}
    \toprule
    \rowfont{\bfseries\sffamily}& & \multicolumn{2}{c}{Plasticine (\si{mm^2})} & & \multicolumn{2}{c}{\name (\si{mm^2})} \\\cmidrule(r){3-4} \cmidrule(l){6-7}
                             \rowfont{\bfseries\sffamily} &  & Each  & Total   &  & Each  & Total   \\ \midrule
    Compute Unit                                            &  & 0.401 & 80.2 &  & 0.423 & 84.7 \\ 
    \tabin\emph{Scanner}                              &  &       &        &  & 4.7\%  &        \\
    \tabin\emph{Format Conv.}                              &  &       &        &  & 0.5\%  &        \\
    \midrule                                             
    Memory Unit                                             &  & 0.199 & 39.7 &  & 0.251 & 50.2 \\ 
    \tabin\emph{Func. Units}                              &  &       &        &  & 4.5\% &        \\
    \tabin\emph{Allocator}                                &  &       &        &  & 0.8\% &        \\
    \midrule                                             
    DRAM AG                                             &  & 0.030 & 2.4  &  & 0.087 & 6.9  \\ 
    \tabin\emph{Func. Units}                              &  &       &        &  & 13.8\%  &        \\
    \tabin\emph{Decompressor}                             &  &       &        &  & 6.0\%  &        \\
    \midrule                                             
    Shuffle Networks                                       &  &       &        &  & 1.064 & 6.4  \\ 
    On-Chip Net                                           &  & 0.075 & 36.3 &  & 0.075 & 36.3 \\
    \midrule                                             
    Total Area (\si{mm^2})                                            &  &       & 158.6    &  &       & 184.5    \\
    Design Power  (\si{W})                                 &  &       & 155   &  &       & 174    \\
    \bottomrule
  \end{tabu}
\end{table}

To estimate \name's hardware requirements, we synthesize Plasticine and the added \name units with Synopsys Design Compiler and the \SI{15}{nm} FreePDK15 predictive library~\cite{freepdk15} at \SI{1.6}{GHz}.
Because FreePDK15 lacks a memory compiler, we scale SRAM area from a \SI{28}{nm} industrial memory compiler; this scaling is represented equally in the Plasticine and \name results.
To simplify application mapping (important for getting good on-chip network performance~\cite{prabhakar2017plasticine}), we matched our baseline's homogeneous compute layout with one scanner per compute tile and one scheduler per memory tile.

Overall, \name is \morearea\% larger than Plasticine and consumes \morepower\% more on-die power (\Cref{tab:area})---with \name's speedup, this would yield much lower energy consumption.
\name's critical path is in the CU's compute pipeline, so it has the same clock frequency and dense performance as Plasticine.
A detailed area comparison of \name and the V100 is challenging because the GPU has both additional hardware (DRAM interfaces, IO, etc.) and more efficient physical design.
However, the significant difference (\SI{\totsize}{mm^2} vs. \SI{815}{mm^2}) points to \name beating the V100 by a wide margin.

Finally, our area and power overheads are shown for a homogeneous design: every CU, MU, and AG is sparsity-enabled, which demonstrates the best sparse performance achievable.
However, if sparse algorithms were less important than dense ones, a designer could provision a fraction of the sparse logic (e.g., 50\% of CUs and MUs with sparsity).
This would halve peak sparse performance while linearly decreasing the area and power overhead.
\begin{table}
  \centering
  \footnotesize
  \caption{\name's sensitivity to SpMU architecture. The ideal case has no SRAM bank conflicts.}
  \label{tab:spmu}
  \pgfplotstabletypeset[
    col sep=comma,
    numeric type,
    column type={r},
    clear infinite,
    assume math mode,
    fixed,
    fixed zerofill,
    precision=2,
    display columns/0/.style={string type,
                              column name={},
                              preproc cell content/.code={}},
    display columns/1/.style={ column name={Ideal}},
    display columns/2/.style={ column name={Hash}},
    display columns/3/.style={ column name={Lin.}},
    display columns/4/.style={ column name={Hash}},
    display columns/5/.style={ column name={Lin.}},
    display columns/6/.style={ column name={Hash}},
    display columns/7/.style={ column name={Lin.}},
    assign column name/.style={
      /pgfplots/table/column name={\sffamily\textbf{#1}}
    },
    every head row/.style={before row={\toprule&&\multicolumn{2}{c}{\sffamily\textbf{\name}}&\multicolumn{2}{c}{\sffamily\textbf{Weak Alloc}}&\multicolumn{2}{c}{\sffamily\textbf{Arb.}}\\\cmidrule(r){3-4}\cmidrule(lr){5-6}\cmidrule(l){7-8}}, after row=\midrule},
    every last row/.style={before row=\midrule, after row=\bottomrule},
    ]{csv/spmu.csvT}
\end{table}

\subsection{Sensitivity Studies}
In \Cref{sec:uarch}, we performed several sensitivity studies using synthesis and microbenchmark simulation to quantify the trade-offs involved in individual units.
Here, we detail broader studies performed using full applications.

\paragraph*{Resources}
One reason that \name was designed as an extension to Plasticine is to \emph{scale up} to large chips  and high memory bandwidth.
\Cref{fig:system}\added{b} demonstrates \name's area scaling as outer-parallelization is varied---nearly linear scaling implies that \name could be extended to even larger die sizes if memory bandwidth were available.
Similarly, the most data-parallel \name applications (SpMV, PR) are memory-bound even with \SI{900}{GB/s} HBM2 memory bandwidth (\added{\Cref{fig:system}a}); even less parallel applications (BFS, SSSP) still benefit from up to \SI{500}{GB/s}.
PREdge and COO see the best compression speedups (\added{\Cref{fig:system}c}) because they load two pointers for every data element, with repeated (i.e., highly compressible) source-node pointers.
\begin{table}
  \centering
  \footnotesize
  \caption{The impact of SpMU ordering modes. Runtimes are normalized to the full-reordering case.}
  \label{tab:ordering_apps}
  \begin{filecontents*}{csv/spmu_mode.csv}
key,CSR,COO,CSC,Conv,BiCGStab,gmean
Capstan,1.0,1.0,1.0,1.0,1.0,1.0
Address Ordered,1.2692939020225378,1.2654211354333114,1.114584289198692,1.6822097398954878,1.4769000513376365,1.3478088033232383
Ordered,1.3463274680354178,4.184469875353148,1.1485409410873775,2.0710534894102515,1.6230245245502841,1.851359701167101
\end{filecontents*}
\pgfplotstabletypeset[
    col sep=comma,
    numeric type,
    column type={r},
    clear infinite,
    assume math mode,
    fixed,
    fixed zerofill,
    precision=2,
    display columns/0/.style={string type,
                              column name={},
                              preproc cell content/.code={}},
    assign column name/.style={
      /pgfplots/table/column name={\sffamily\textbf{#1}}
    },
    every head row/.style={before row=\toprule, after row=\midrule},
    every last row/.style={after row=\bottomrule},
    ]{csv/spmu_mode.csv}
\end{table}

\begin{table}
  \centering
  \footnotesize
  \caption{\name's sensitivity to the merge network. Runtimes are normalized to our primary design point, Mrg-1.}
  \label{tab:merge}
  \pgfplotstabletypeset[
    col sep=comma,
    numeric type,
    column type={r},
    clear infinite,
    assume math mode,
    fixed,
    fixed zerofill,
    precision=2,
    columns={key,DDR4-NoShuf,HBM2-NoShuf,Mrg-0,Mrg-1,Mrg-16},
    display columns/0/.style={string type,
                              column name={},
                              preproc cell content/.code={}},
    display columns/1/.style={column name={None}},
    display columns/2/.style={column name={None}},
    assign column name/.style={
      /pgfplots/table/column name={\sffamily\textbf{#1}}
    },
    every head row/.style={before row={\toprule
    & \multicolumn{1}{c}{\sffamily\bfseries DDR4} & \multicolumn{4}{c}{\sffamily\bfseries HBM2E}\\\cmidrule(lr){2-2}\cmidrule(lr){3-6}}, after row=\midrule},
    every last row/.style={after row=\bottomrule},
    ]{csv/shuf.csvT}
\end{table}

\begin{table*}
  \footnotesize
  \centering
  \caption{Runtimes normalized to the fastest \name-HBM2E version of each application. Bolded points are used for geomeans; not all baselines support all application variants.}
  \label{tab:perf}
  \begin{filecontents*}{csv/main.csv}
key,CSR,COO,CSC,Conv,Pull,Edge,BFS,SSSP,M+M,SpMSpM,BiCGStab,gmean
Capstan (Ideal Net \& Mem),0.8277586401001291,1.2077407406313507,0.8120847443795614,0.9520265947179303,0.7935883529778844,1.0607341965444306,0.654399460317569,0.7320264346282948,0.8622428651500897,0.8820189274447949,0.9441386784938293,0.8232748336333603
Capstan (HBM2E),1.2529528790796176,1.670000770772425,1.0,1.0,1.0,1.3315564223462517,1.0,1.0,1.0,1.0,1.0,1.0
Capstan (HBM2),1.7821517814973373,2.255164312186128,1.2741291665495655,1.0066309006484269,1.3721045216769876,1.72667196758196,1.2804266967621698,1.203855759358624,1.3508633050026317,1.5286492687123603,1.1872782289445385,1.2672403156262457
Capstan (DDR4),18.15670330626116,21.942029094403775,10.494985090150372,1.5264111348409055,12.079794319270615,14.002920735327653,5.241526144694876,3.8898043832886167,8.2019352234448,6.888328075709779,13.434194991441927,6.449744737850318
Plasticine (HBM2E),17.03678667001163,184.16172360384886,365.09087090052225,nan,8.483701051498812,nan,nan,nan,nan,nan,7.566555006149937,10.302842918849489
V100 GPU,6.162953336989674,nan,nan,119.39068400104802,8.681543527235782,nan,31.639867365693835,13.586557836918553,12.253138979145916,41.78996271866935,22.18806643820487,20.500638874839694
128-Thread CPU,67.85662240369072,640.3059434641361,485.6446602537433,99.86326928587718,52.90798294661315,nan,62.28628236077217,68.28920288438322,73.90448507609291,2254.0866073989105,143.0256721543559,117.50054569069617
\end{filecontents*}
\pgfplotstabletypeset[
    col sep=comma,
    numeric type,
    column type={r},
    clear infinite,
    assume math mode,
    fixed,
    fixed zerofill,
    precision=2,
    every row 0 column 3/.style={normpoint},
    every row 0 column 4/.style={normpoint},
    every row 0 column 5/.style={normpoint},
    every row 0 column 7/.style={normpoint},
    every row 0 column 8/.style={normpoint},
    every row 0 column 9/.style={normpoint},
    every row 0 column 10/.style={normpoint},
    every row 0 column 11/.style={normpoint},
    every row 1 column 3/.style={normpoint},
    every row 1 column 4/.style={normpoint},
    every row 1 column 5/.style={normpoint},
    every row 1 column 7/.style={normpoint},
    every row 1 column 8/.style={normpoint},
    every row 1 column 9/.style={normpoint},
    every row 1 column 10/.style={normpoint},
    every row 1 column 11/.style={normpoint},
    every row 2 column 3/.style={normpoint},
    every row 2 column 4/.style={normpoint},
    every row 2 column 5/.style={normpoint},
    every row 2 column 7/.style={normpoint},
    every row 2 column 8/.style={normpoint},
    every row 2 column 9/.style={normpoint},
    every row 2 column 10/.style={normpoint},
    every row 2 column 11/.style={normpoint},
    every row 3 column 3/.style={normpoint},
    every row 3 column 4/.style={normpoint},
    every row 3 column 5/.style={normpoint},
    every row 3 column 7/.style={normpoint},
    every row 3 column 8/.style={normpoint},
    every row 3 column 9/.style={normpoint},
    every row 3 column 10/.style={normpoint},
    every row 3 column 11/.style={normpoint},
    every row 5 column 1/.style={normpoint},
    every row 5 column 4/.style={normpoint},
    every row 5 column 5/.style={normpoint},
    every row 5 column 7/.style={normpoint},
    every row 5 column 8/.style={normpoint},
    every row 5 column 9/.style={normpoint},
    every row 5 column 10/.style={normpoint},
    every row 5 column 11/.style={normpoint},
    every row 4 column 1/.style={normpoint},
    every row 4 column 5/.style={normpoint},
    every row 4 column 11/.style={normpoint},
    every row 6 column 1/.style={normpoint},
    every row 6 column 4/.style={normpoint},
    every row 6 column 5/.style={normpoint},
    every row 6 column 7/.style={normpoint},
    every row 6 column 8/.style={normpoint},
    every row 6 column 9/.style={normpoint},
    every row 6 column 10/.style={normpoint},
    every row 6 column 11/.style={normpoint},
    display columns/0/.style={string type,
                              column name={},
                              preproc cell content/.code={}},
    assign column name/.style={
      /pgfplots/table/column name={\sffamily\textbf{#1}}
    },
    every head row/.style={before row={\toprule&\multicolumn{3}{c}{\sffamily\textbf{SpMV}}&&\multicolumn{2}{c}{\sffamily\textbf{PageRank}}&&&&\\\cmidrule{2-4}\cmidrule{6-7}}, after row=\midrule},
    every last row/.style={after row=\bottomrule},
    ]{csv/main.csv}
\end{table*}

\begin{figure*}
  \centering
\pgfplotstableread[col sep=comma]{csv/scan.csv}\tabscan
\pgfplotstableread[col sep=comma]{csv/data_scan.csv}\tabdatascan
\pgfplotstableread[col sep=comma]{csv/vec_scan.csv}\tabvecscan
  \begin{tikzpicture}
    \begin{groupplot}[
scanheight,
groupbase,
    legend style={
      legend columns=4,
      at={(0.5, 1.05)},
      anchor=south,
      draw=none,
      /tikz/every even column/.append style={column sep=0mm},
    },
  ylabel={Slowdown},
    set layers,
]
        \nextgroupplot[
  ylabel={Slowdown},
  xlabel={\added{(a)} Bits Scanned per Cycle},
  ymode=log,
  xmode=log,
  legend entries={BFS,SSSP,M+M,SpMSpM},
   mark options={draw opacity=0, scale=0.5},
    xmin=1,
    xmax=512,
    xtick={1,4,16,64,512},
    xticklabels={1,4,16,64,512},
    ymin=1.0,
    ymax=3,
    ytick={1,2,3},
    yticklabels={1,2,3},
]
  \addplot table[x=key, y=BFS]  {\tabscan};
  \addplot table[x=key, y=SSSP]  {\tabscan};
  \addplot table[x=key, y=MMAddTree]  {\tabscan};
  \addplot table[x=key, y=RowGEMM]  {\tabscan};
        \nextgroupplot[
  xlabel={\added{(b)} Data Scanned per Cycle},
  xmode=log,
  legend entries={CSC,Conv},
   mark options={draw opacity=0, scale=0.5},
    xmin=1,
    xmax=16,
    ymin=1.0,
    ymax=1.2,
    xtick={1,2,4,8,16},
    xticklabels={1,2,4,8,16},
    ytick={1,1.05,1.10,1.15},
    yticklabels={1,1.05,1.10,1.15},
]
  \addplot table[x=key, y=CSC]  {\tabdatascan};
  \addplot table[x=key, y=Conv]  {\tabdatascan};
        \nextgroupplot[
  ylabel={Slowdown},
  xlabel={\added{(c)} Scan Output Vectorization},
  ymode=log,
  xmode=log,
  legend entries={M+M, SpMSpM},
   mark options={draw opacity=0, scale=0.5},
    xmin=1,
    xmax=16,
    ymin=1.0,
    ymax=3,
    xtick={1,2,4,8,16},
    xticklabels={1,2,4,8,16},
    ytick={1,2,3},
    yticklabels={1,2,3},
]
  \addplot table[x=key, y=MMAddTree]  {\tabvecscan};
  \addplot table[x=key, y=RowGEMM]  {\tabvecscan};
    \end{groupplot}

  \end{tikzpicture}
    \caption{\name's sensitivity to scanner width, relative to a maximal (512 input, 16 output) scanner.}
    \label{fig:scancomp}
\end{figure*}
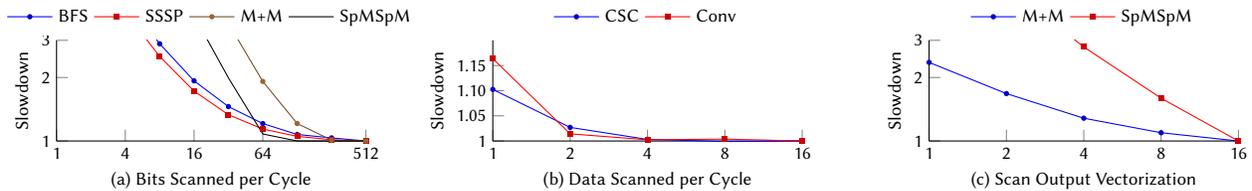

\paragraph*{On-Chip Memories}
In addition to system-level evaluations, we performed several sensitivity studies to capture the impact of \name's allocated sparse memories and cross-partition communication for outer-parallelized apps.
If \name's memories were arbitrated (executing accesses from one vector at at time), it would be \arbslow\% slower on average and up to \arbslowmax\% slower for certain applications (\Cref{tab:spmu}), even with address hashing.
As we previously demonstrated, using a multi-iteration, multi-priority allocator further improves theoretical throughput, and this carries forward to a system-level speedup---an average of 15\% and peak of 39\%.
The allocator is small (\Cref{tab:area}), so these speedups are effectively free, just from using the crossbar and memory banks more efficiently.
\name's allocated design is also only \slowvsideal\% slower than an ideal SpMU that does not model any bank conflicts.
Finally, \Cref{tab:ordering_apps} confirms, at a system level, the microbenchmark results in \Cref{fig:order}.

The shuffle network (\Cref{tab:merge}) improves performance by keeping cross-tile accesses on-chip; the speedup is more pronounced for DDR4, which has less bandwidth.
 As mentioned in \Cref{sec:uarchshuffle}, the shuffle network is not a full crossbar, which would be too expensive.
Instead, we shift input lanes in each vector by up to one position (Mrg-1) to increase the merge success rate.
Shifting arbitrarily (full crossbar, Mrg-16) provides minimal benefits, and not shifting lanes at all (Mrg-0) slows down applications with cross-partition communication.
Data partitioning reduces cross-partition communication for graph applications, but convolution uses the shuffle network to avoid a separate halo-exchange pass. 
For convolutions with $3\times3$ kernels, Mrg-0 is up to 15\% slower.

\paragraph*{Scanners}
Finally, we investigated scanners' impact on performance (\Cref{fig:scancomp}).
Bit scanners---which traverse more zeros---are sensitive to scanner length, with a massive slowdown for the scalar case (a single bit).
Even scanning 128 bits would slow M+M by 21\%, so we scan 256 bits per cycle.
Shorter data scanners are less impactful, with a peak slowdown of only 16\% on Conv.
However, data scanners are small, so we use a data scanner with 16 inputs.

Scanner output vectorization is also important, with impact varying based on the application.
Only outputting eight elements per cycle has a significant performance impact on SpMSpM, because its datasets are relatively dense; for M+M, which is evaluated with sparser matrices, the gains from added vectorization are lower.
However, even very sparse matrices have denser sections (frequently, near the diagonal); vectorization helps run these faster.

\subsection{Performance}

\Cref{tab:perf} shows \name's performance compared to several optimized baseline designs. 
CPU baselines are TACO~\cite{taco} for sparse linear algebra and GraphIt~\cite{zhang2018graphit}, both using 128 threads on a four-socket Xeon E7-8890~v3~\cite{xeon8890v3}; GPU baselines are cuSparse~\cite{cuSparse} and Gunrock~\cite{wang2016gunrock} run on a V100 GPU~\cite{v100datasheet}.
\name outperforms the CPU and GPU baselines by a large margin, with the largest performance increases for applications relying heavily on sparse memory accesses. 
Several \name features, including cross-tile sparse updates (Conv), sparse DRAM updates (PREdge), and sparse iteration (BFS, SSSP, M+M, and SpMSpM) can not be mapped efficiently to Plasticine, so only some applications have Plasticine baselines.

To demonstrate the benefits of streaming kernel fusion, we implemented a stabilized Biconjugate Gradient solver (BiCGStab)~\cite{van1992bi}; this is a linear least squares solver that combines sparse matrix-vector multiplication and dense dot products.
The CPU and GPU baselines implement BiCGStab using sparse and dense kernels; the inter-kernel overhead causes up to a 3\x slowdown relative to sparse SpMV alone.
However, \name (and Plasticine) can fuse these kernels into a streaming pipeline, which lowers memory bandwidth requirements \emph{and} the latency of each iteration.
\name's Plasticine-relative speedup is lower on BiCGStab because they have the same dense performance.

\paragraph*{ASIC Comparison}
\label{sec:eval:asic}
\Cref{tab:asiccomp} shows \name's performance compared against recently-proposed ASICs.
The EIE and SCNN comparisons are run against an ideal (i.e., ignoring network delays, bank conflicts, and load/store time) model of each baseline.
When comparing against EIE, we exclude \name's network latency and load/store time to provide a comparison of compute throughput between the two architectures.
For SCNN, we use the manually mapped non-positional variant described earlier in this section.
The Graphicionado\footnote{Graphicionado does not report area; we expect that the large eDRAM scratchpad would dominate its area.} comparison is run against published edge-processing rates on the flickr and fb~\cite{wilson2009user} datasets using DDR4 memory for \name, providing a more accurate result. 
Also, we use BFS and SSSP variants that do not write back-pointers for a fairer Graphicionado comparison.
Mat\-Raptor is compared against its highest demonstrated throughput (\SI{10}{GOP/s}).
Because the Graphicionado and MatRaptor baselines use detailed simulations, we include load and store time for the \name comparison.
\begin{table}
  \centering
  \caption{\name compared to recent accelerators.}
  \label{tab:asiccomp}
  \footnotesize
  \begin{tabu}{clrrr}
    \toprule
    \rowfont{\sffamily\bfseries}                                                  & & \multicolumn{2}{c}{Speedup (\x)} & \\ \cmidrule{3-4}
    \rowfont{\sffamily\bfseries}                                                  & App    & \SI{1.6}{GHz}& \SI{1}{GHz} & Ref. Area                           \\ \midrule
    \sffamily\bfseries EIE \cite{han2016eie}                                      & CSC    & \ubold{0.53}        &0.40& \SI{64}{mm^2}/\SI{28}{nm}           \\ \midrule
    \sffamily\bfseries SCNN \cite{parashar2017scnn}                               & Conv   & \ubold{1.40}        &0.87& \SI{7.9}{mm^2}/\SI{16}{nm}          \\ \midrule
    \multirow{3}{*}{\sffamily\bfseries Graphicionado \cite{ham2016graphicionado}} & PR     & \ubold{1.08}        &0.97& \multirow{3}{*}{\SI{64}{MiB} eDRAM} \\
                                                                                  & BFS    & \ubold{2.10}        &2.06&                                     \\
                                                                                  & SSSP   & \ubold{1.13}        &1.03&                                     \\ \midrule
    \sffamily\bfseries MatRaptor \cite{srivastava2020matraptor}                   & SpMSpM & \ubold{17.96}       &12.22 & \SI{2.26}{mm^2}/\SI{28}{nm}         \\
    \bottomrule
  \end{tabu}
\end{table}

\name starts with the advantage of a higher core clock (\SI{1.6}{GHz}) than most accelerators, due to aggressive pipelining in the CU and SpMU.
A faster clock provides the biggest advantage for convolution and matrix-matrix multiply, which are almost entirely on-chip; the graph algorithms see little benefit because they are heavily DRAM-limited.
CSC is an intermediate case, having significant on-chip processing interspersed with DRAM loads of matrix data.

\name under-performs EIE for two key reasons: EIE stores model weights entirely on-chip, while \name fetches them from HBM, and EIE has more scalar tiles, which do not suffer from under-vectorization.
Vectorization can provide significant benefits, though, by increasing peak throughput. 
\name can process up to 128 elements per cycle---16\x  more than MatRaptor, which uses eight scalar pipelines.
\name and Graphicionado are both DRAM-limited; this comparison shows that \name does not rely solely on DRAM bandwidth for performance.

SCNN is designed using a 2-D multiplier array that processes four activations and four weights per cycle; for layers with few activations, 75\% of this array is unused.
However, for layers with more activations and fewer weights, the shorter vector lengths can increase utilization.
\name's relatively large on-chip memory also provides a slight performance benefit: SCNN is forced to tile its outputs, which limits the amount of available weight parallelism and forces multiple iterations.

\begin{figure}
    \pgfplotstableread[col sep=comma]{csv/breakdown.csv}\tab
\tikzstyle{labelstyle}=[font=\footnotesize, rotate=45,anchor=north east]
\begin{tikzpicture}
\begin{axis}[
    globplot,
    height=1.4in,
    font=\footnotesize,
    enlarge x limits = 0.05,
    ybar stacked,
    width=\linewidth,
    bar width=4.5pt,
      ymin=0.0,
      xticklabel style={anchor=north east},
      ylabel={Time (\%)},
    xtick=\empty,
  legend entries={Active,Scan,Load/Store,Vector Length,Imbalance,Network,SRAM,DRAM},
    legend style={ at={(0.5,1.0)}, 
    legend columns = 4,
anchor=south
},
clip=false,
    cycle list name=stallcycle,
    cycle list shift=1,
]
  \addplot table[x=xaxis, meta=app, y expr=100*\thisrow{good}]  {\tab};
  \addplot table[x=xaxis, meta=app, y expr=100*\thisrow{scan}]  {\tab};
  \addplot table[x=xaxis, meta=app, y expr=100*\thisrow{xfer}]  {\tab};
  \addplot table[x=xaxis, meta=app, y expr=100*\thisrow{vec}]  {\tab};
  \addplot table[x=xaxis, meta=app, y expr=100*\thisrow{imb}]  {\tab};
  \addplot table[x=xaxis, meta=app, y expr=100*(\thisrow{net})]  {\tab};
  \addplot table[x=xaxis, meta=app, y expr=100*\thisrow{bank}]  {\tab};
  \addplot table[x=xaxis, meta=app, y expr=100*\thisrow{mem}]  {\tab};
  \node[labelstyle] at (axis cs:1,0) {CSR};
  \node[labelstyle] at (axis cs:5,0) {COO};
  \node[labelstyle] at (axis cs:9,0) {CSC};
  \node[labelstyle] at (axis cs:13,0) {Conv};
  \node[labelstyle] at (axis cs:17,0) {PRPull};
  \node[labelstyle] at (axis cs:21,0) {PREdge};
  \node[labelstyle] at (axis cs:25,0) {BFS};
  \node[labelstyle] at (axis cs:29,0) {SSSP};
  \node[labelstyle] at (axis cs:33,0) {M+M};
  \node[labelstyle] at (axis cs:37,0) {SpMSpM};

\end{axis}
\end{tikzpicture}
    \vskip-\baselineskip
    \caption{A breakdown of \name's execution time, with three datasets evaluated per application. Datasets are shown here in the same order as in \Cref{tab:lindata}.}
    \label{fig:stalls}
  \end{figure}
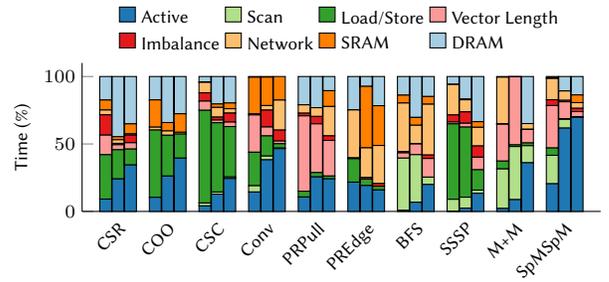

  \paragraph*{Stall Breakdown}
Every \name application has multiple outer-parallel pipelines, each with sixteen vector lanes.
In \Cref{fig:stalls}, we look at limiting factors to identify why every lane is not performing useful work in every cycle.
We start with a synthetic analysis, which tells us how many cycles would be needed if every lane were active in every cycle (Active).
We then look at lanes that are inactive because their associated scanner is processing an all-zero vector (Scan) and lanes that are waiting for data to be loaded from or stored to DRAM (Load/Store). 
For the synthetic analysis, load/store time assumes zero-latency, infinite-bandwidth DRAM.
Next, our synthetic analysis shows lanes that are underused because vectorized loops are too short (Vector Length) or because workload tiling generates unevenly-sized tiles (Imbalance).
We then simulate, adding in on-chip pipelining and network effects (Network), bank conflicts (SRAM), and the Ramulator HBM2E model (DRAM).
By adding these one at at time, we identify the cycles that are lost to each stall source.

The on-chip network has a large impact on BFS and SSSP because they cannot be pipelined between iterations; conversely, SpMSpM can easily be pipelined, which leads to high activity factors.
Load times are over-represented for some applications (COO--CSC) because we assume that all data is loaded from and stored to DRAM and measure end-to-end latency for a single iteration. 
If these kernels were run as part of a pipeline, data could be passed through SRAM multi-buffers~\cite{prabhakar2017plasticine} and efficiency would increase.
Moreover, PRPull and PREdge behave differently. 
PRPull suffers from under-vectorization because many graph vertices have very few in-edges.
However, PREdge suffers from SRAM conflicts on datasets which have a power-law distribution, where some vertices have many in-edges that cannot be coalesced.
Therefore, it is important to be able to choose between pull and edge-based execution.

\section{Related Work}
\paragraph*{Graph Accelerators}
FPGAs have been proposed for graph analytics, including primitives~\cite{zhang2017boosting, zhou2015accelerating, mcgettrick2008fpga, oguntebi2016graphops} and specialized compilers~\cite{nurvitadhi2014graphgen, dai2016fpgp}.
However, FPGAs are \emph{too flexible:} \name provides optimized iteration hardware for common parallel patterns, with better logic density and faster clocks.
Graph accelerator ASICs like Tesseract~\cite{ahn2016scalable,zhang2018graphp}, GraphR~\cite{song2018graphr}, and Graphicionado~\cite{ham2016graphicionado} use memory systems (HMC, ReRAM, and eDRAM) and architectures specialized for graph analytics.
Intel's PIUMA~\cite{aananthakrishnan2020piuma} is similar but CPU-based, with highly multi-threaded cores and gather/scatter offload engines.
Finally, PolyGraph~\cite{dadu2021polygraph}, which adds a task-scheduling core to an RDA for frontier-set graph algorithms, demonstrates the value of switching between graph algorithm variants.
However, separating graph analytics and linear algebra may preclude new applications, like graph neural networks~\cite{zhang2018gnn}. 
Furthermore, having high-level compilers targeting performant hardware for graph analytics will let domain experts focus on optimizing their algorithms instead of forcing them to design hardware or write low-level code.

\paragraph*{Sparse Linear Algebra Accelerators}
Other ASICs have been proposed for sparse linear algebra and deep-learning inference, as summarized by Dave et al.~\cite{dave2020hardware}.
ExTensor~\cite{hegde2019extensor} uses hardware to hierarchically eliminate computations on sparse tensors. 
Gamma~\cite{gamma} and MatRaptor~\cite{srivastava2020matraptor} implement row-product SpMSpM while OuterSPACE~\cite{pal2018outerspace} and SpArch~\cite{zhang2020sparch} use outer-product SpMSpM; \name supports these algorithms (but not SpArch's priority-queue-driven scheduler).
\name adds vectorization with bit-vector sparsity.
These architectures can only process one intersection per compute stream, per cycle, while \name can process up to 16 intersections in a single CU.

Bespoke deep-learning accelerators also limit the available parallelism dimensions: for example, the A100~\cite{choquette2020nvidia} only supports a sparse matrix-matrix multiply with a restrictive two-of-four format.
SCNN~\cite{parashar2017scnn} requires that parallelism be exploited first across data tiles and then across four weights and four activations per cycle.
Cnvlutin~\cite{albericio2016cnvlutin} skips zeros in CNN activations, and Eyeriss~\cite{chen2016eyeriss} and EIE~\cite{han2016eie} skip zeros in activations and weights using CSC.
\name, on the other hand, has generic sparse iteration primitives, flexible inner- and outer-loop parallelism, and automatic pipelining.
\name's allocated scratchpad adds area relative to specialized memories like Cambricon-X's~\cite{zhang2016cambricon} offset indexing.
However, the scratchpad is far more powerful, letting programmers change data formats without re-designing hardware.

\paragraph*{Plasticine \& Spatial}
Plasticine's programs are statically banked so no two lanes access the same memory bank in a cycle; memory duplication and buffering are performed as necessary.
In the worst banking cases (random accesses), each memory only supports one access per cycle, leaving 15 banks inactive.
Plasticine also does not permit read-modify-write (RMW) accesses---for consistent random RMWs, each read must block on the preceding write, introducing multi-cycle bubbles.
This is most visible in COO and CSC SpMV, which rely on \emph{modifying} data.
Furthermore, Plasticine has no sparse iteration support, which limits which programs that can be mapped.
Finally, Gorgon~\cite{vilim2020gorgon}, which is based on Plasticine, can perform joins but lacks support for random memory accesses and updates.

\paragraph*{Other RDAs}
Reconfigurable architectures like Dy\-SER~\cite{govindaraju2012dyser}, Wave\-Scalar~\cite{swanson2003wavescalar}, Pipe\-Rench~\cite{goldstein1999piperench}, and TRIPS~\cite{burger2004scaling} provide support for efficient computation.
However, these are designed to execute small, scalar program fragments, which prevents the loop-unrolling parallelism that Plasticine and \name use.
The MIT Raw~\cite{taylor2001raw} processor provides sparse accesses, but relies on the compiler to ensure memory consistency; its sparse accesses are also not vectorized.
Compared to these designs, \name brings Plasticine's dense performance improvements to a wider range of sparse problems.

Dadu et al.~\cite{dadu2019towards} have recently proposed the SPU, an architecture with tiles combining a sparse memory with an RDA fabric.
Each SPU has both a highly-banked scratchpad optimized for sparse operations and a linear scratchpad optimized for dense operations.
However, the SPU uses a 16\x{}32 crossbar that only exposes 50\% of bank bandwidth---a lower limit than \name's average performance---and its stream-join programming model prevents vectorized sparse iteration.
\added{The stream-join model exposes an \emph{arbitrary} compare-dequeue abstraction, where the results of one comparison change the dequeue decision and thus the arguments to the next comparison.
Therefore, vectorizing stream-join would require packing 16 sequential comparisons into a single cycle, which scales poorly to high frequencies.}
Finally, the SPU's alias-free indirection corresponds roughly to \name's unordered memory mode, which precludes applications that require stricter memory ordering.

\paragraph*{Optimizing Accelerator Memories}
\added{Other works propose logical primitives for transferring and packing data into accelerator scratchpads.}
Buffets~\cite{pellauer2019buffet} coordinate loading and sharing tiles of data between levels of the memory hierarchy, and Stash~\cite{komuravelli2015stash} provides cache-like functionality without the overhead of tag checks.
CoRAM~\cite{chung2011coram} optimizes memory transfers to and from an FPGA's fabric using a programmable, DMA-like interface.
\name's SpMU is orthogonal to these approaches: for example, Buffets could be used instead of Spatial to coordinate bulk data transfers between the SpMU and DRAM.
Alternately, \name's allocated pipeline could be added to a Buffet or Stash scratchpad, which would simultaneously provide random-access performance and data coordination for dedicated accelerators.
Memory compression has also been used before, including for GPUs~\cite{krajcevski2016,shim2004,heckbert1982} and sparse matrices~\cite{kanellopoulos2019}; \name's approach is similar to GPUs' delta-compression but specialized for sparse pointers.

\paragraph*{Allocation}
Allocators have also been explored in networking for crossbar configuration~\cite{dally2004principles,becker2009allocator}.
Network allocators optimize for network metrics, including iSLIP~\cite{mckeown1999islip} and PIM~\cite{anderson1993high} for fairness and age-based allocation for QoS~\cite{jiang2011performance}.
\name introduces these concepts to memory bank scheduling, evaluating allocation's benefit and trade-offs in a new context.

\section{Conclusion}
In this paper, we introduce \name, a reconfigurable, vectorized, streaming, scalable sparse accelerator that also supports dense data analytics.
\name unifies applications and hardware with sparse iteration modes: these permit algorithmic optimizations \emph{and} hardware optimizations like vectorization.
\name's novel bank-allocation scheme also provides high random throughput (80\%) to distributed sparse memories---more than double that of an arbitrated baseline.

With only \morearea\% more area and \morepower\% more power, \name brings new applications to Plasticine and runs existing ones 7.6\x to 365\x faster. 
\name is also significantly faster than CPU, GPU, and several accelerator baselines; the only baseline with better performance is EIE~\cite{han2016eie} because it is able to store models entirely on-chip, which is impractical for a general-purpose accelerator.
We believe that \name will drive future research into sparse hardware---it demonstrates that vectorized RDAs can run sparse applications and that using an explicit, limited set of sparse iteration primitives lets hardware extract more parallelism.

\begin{acks}
We would like to thank Muhammad Shahbaz and the anonymous reviewers for their feedback on this paper.
This work has been supported by Stanford's DAWN Lab funders, a gift from Facebook, and a Stanford Graduate Fellowship.
This material is based on research sponsored by Defense Advanced Research Projects Agency (DARPA) under agreements numbered FA8750-17-2-0095 and FA8750-14-2-0240 and Air Force Research Laboratory (AFRL) and DARPA under agreement number FA8650-18-2-7865.
The U.S. Government is authorized to reproduce and distribute reprints for Governmental purposes notwithstanding any copyright notation thereon.
The views and conclusions contained herein are those of the authors and should not be interpreted as necessarily representing the official policies or endorsements, either express or implied, of AFRL and DARPA or the U.S. Government.
This material is also supported by NSF awards 1937301, 2028602, CCF-1563078, and 1563113.
\end{acks}

\bibliographystyle{ACM-Reference-Format}
\bibliography{paper}


\begin{thebibliography}{80}


\ifx \showCODEN    \undefined \def \showCODEN     #1{\unskip}     \fi
\ifx \showDOI      \undefined \def \showDOI       #1{#1}\fi
\ifx \showISBNx    \undefined \def \showISBNx     #1{\unskip}     \fi
\ifx \showISBNxiii \undefined \def \showISBNxiii  #1{\unskip}     \fi
\ifx \showISSN     \undefined \def \showISSN      #1{\unskip}     \fi
\ifx \showLCCN     \undefined \def \showLCCN      #1{\unskip}     \fi
\ifx \shownote     \undefined \def \shownote      #1{#1}          \fi
\ifx \showarticletitle \undefined \def \showarticletitle #1{#1}   \fi
\ifx \showURL      \undefined \def \showURL       {\relax}        \fi
\providecommand\bibfield[2]{#2}
\providecommand\bibinfo[2]{#2}
\providecommand\natexlab[1]{#1}
\providecommand\showeprint[2][]{arXiv:#2}

\bibitem[\protect\citeauthoryear{Aananthakrishnan, Ahmed, Cave, Cintra, Demir,
  Bois, Eyerman, Fryman, Ganev, Heirman, et~al\mbox{.}}{Aananthakrishnan
  et~al\mbox{.}}{2020}]%
        {aananthakrishnan2020piuma}
\bibfield{author}{\bibinfo{person}{Sriram Aananthakrishnan},
  \bibinfo{person}{Nesreen~K Ahmed}, \bibinfo{person}{Vincent Cave},
  \bibinfo{person}{Marcelo Cintra}, \bibinfo{person}{Yigit Demir},
  \bibinfo{person}{Kristof~Du Bois}, \bibinfo{person}{Stijn Eyerman},
  \bibinfo{person}{Joshua~B Fryman}, \bibinfo{person}{Ivan Ganev},
  \bibinfo{person}{Wim Heirman}, {et~al\mbox{.}}}
  \bibinfo{year}{2020}\natexlab{}.
\newblock \showarticletitle{{PIUMA}: Programmable Integrated Unified Memory
  Architecture}.
\newblock \bibinfo{journal}{\emph{arXiv preprint arXiv:2010.06277}}
  (\bibinfo{year}{2020}).
\newblock


\bibitem[\protect\citeauthoryear{Ahn, Hong, Yoo, Mutlu, and Choi}{Ahn
  et~al\mbox{.}}{2016}]%
        {ahn2016scalable}
\bibfield{author}{\bibinfo{person}{Junwhan Ahn}, \bibinfo{person}{Sungpack
  Hong}, \bibinfo{person}{Sungjoo Yoo}, \bibinfo{person}{Onur Mutlu}, {and}
  \bibinfo{person}{Kiyoung Choi}.} \bibinfo{year}{2016}\natexlab{}.
\newblock \showarticletitle{A Scalable Processing-in-Memory Accelerator for
  Parallel Graph Processing}.
\newblock \bibinfo{journal}{\emph{ACM SIGARCH Computer Architecture News}}
  \bibinfo{volume}{43}, \bibinfo{number}{3} (\bibinfo{year}{2016}),
  \bibinfo{pages}{105--117}.
\newblock


\bibitem[\protect\citeauthoryear{Albericio, Judd, Hetherington, Aamodt, Jerger,
  and Moshovos}{Albericio et~al\mbox{.}}{2016}]%
        {albericio2016cnvlutin}
\bibfield{author}{\bibinfo{person}{J Albericio}, \bibinfo{person}{P Judd},
  \bibinfo{person}{T Hetherington}, \bibinfo{person}{T Aamodt},
  \bibinfo{person}{N Jerger}, {and} \bibinfo{person}{A Moshovos}.}
  \bibinfo{year}{2016}\natexlab{}.
\newblock \showarticletitle{Cnvlutin: Zero-Neuron-Free Deep Convolutional
  Neural Network Computing}. In \bibinfo{booktitle}{\emph{Proceedings of
  ISCA-43}}.
\newblock


\bibitem[\protect\citeauthoryear{Anderson, Owicki, Saxe, and Thacker}{Anderson
  et~al\mbox{.}}{1993}]%
        {anderson1993high}
\bibfield{author}{\bibinfo{person}{Thomas~E. Anderson},
  \bibinfo{person}{Susan~S. Owicki}, \bibinfo{person}{James~B. Saxe}, {and}
  \bibinfo{person}{Charles~P. Thacker}.} \bibinfo{year}{1993}\natexlab{}.
\newblock \showarticletitle{High-Speed Switch Scheduling for Local-Area
  Networks}.
\newblock \bibinfo{journal}{\emph{ACM Trans. Comput. Syst.}}
  \bibinfo{volume}{11}, \bibinfo{number}{4} (\bibinfo{date}{Nov.}
  \bibinfo{year}{1993}), \bibinfo{pages}{319–352}.
\newblock
\showISSN{0734-2071}
\urldef\tempurl%
\url{https://doi.org/10.1145/161541.161736}
\showDOI{\tempurl}


\bibitem[\protect\citeauthoryear{Barham and Isard}{Barham and Isard}{2019}]%
        {stuckinarut}
\bibfield{author}{\bibinfo{person}{Paul Barham} {and} \bibinfo{person}{Michael
  Isard}.} \bibinfo{year}{2019}\natexlab{}.
\newblock \showarticletitle{Machine Learning Systems Are Stuck in a Rut}. In
  \bibinfo{booktitle}{\emph{Proceedings of the Workshop on Hot Topics in
  Operating Systems}} (Bertinoro, Italy) \emph{(\bibinfo{series}{HotOS '19})}.
  \bibinfo{publisher}{Association for Computing Machinery},
  \bibinfo{address}{New York, NY, USA}, \bibinfo{pages}{177–183}.
\newblock
\showISBNx{9781450367271}
\urldef\tempurl%
\url{https://doi.org/10.1145/3317550.3321441}
\showDOI{\tempurl}


\bibitem[\protect\citeauthoryear{Becker and Dally}{Becker and Dally}{2009}]%
        {becker2009allocator}
\bibfield{author}{\bibinfo{person}{Daniel~U Becker} {and}
  \bibinfo{person}{William~J Dally}.} \bibinfo{year}{2009}\natexlab{}.
\newblock \showarticletitle{Allocator Implementations for Network-on-Chip
  Routers}. In \bibinfo{booktitle}{\emph{Proceedings of the Conference on High
  Performance Computing Networking, Storage and Analysis}}. IEEE,
  \bibinfo{pages}{1--12}.
\newblock


\bibitem[\protect\citeauthoryear{Burger, Keckler, McKinley, Dahlin, John, Lin,
  Moore, Burrill, McDonald, and Yoder}{Burger et~al\mbox{.}}{2004}]%
        {burger2004scaling}
\bibfield{author}{\bibinfo{person}{Doug Burger}, \bibinfo{person}{Stephen~W
  Keckler}, \bibinfo{person}{Kathryn~S McKinley}, \bibinfo{person}{Mike
  Dahlin}, \bibinfo{person}{Lizy~K John}, \bibinfo{person}{Calvin Lin},
  \bibinfo{person}{Charles~R Moore}, \bibinfo{person}{James Burrill},
  \bibinfo{person}{Robert~G McDonald}, {and} \bibinfo{person}{William Yoder}.}
  \bibinfo{year}{2004}\natexlab{}.
\newblock \showarticletitle{Scaling to the End of Silicon with {EDGE}
  Architectures}.
\newblock \bibinfo{journal}{\emph{Computer}} \bibinfo{volume}{37},
  \bibinfo{number}{7} (\bibinfo{year}{2004}), \bibinfo{pages}{44--55}.
\newblock


\bibitem[\protect\citeauthoryear{Cerebras}{Cerebras}{2020}]%
        {cerebras}
\bibfield{author}{\bibinfo{person}{Cerebras}.} \bibinfo{year}{2020}\natexlab{}.
\newblock \bibinfo{title}{The {Cerebras} {CS-1} Product Overview}.
\newblock
\newblock
\urldef\tempurl%
\url{https://secureservercdn.net/192.169.220.245/a7b.fcb.myftpupload.com/wp-content/uploads/2020/01/The-Cerebras-CS-1-Product-Overview-rev20200112.pdf}
\showURL{%
\tempurl}


\bibitem[\protect\citeauthoryear{{Chen, Yu-Hsin and Krishna, Tushar and Emer,
  Joel and Sze, Vivienne}}{{Chen, Yu-Hsin and Krishna, Tushar and Emer, Joel
  and Sze, Vivienne}}{2016}]%
        {chen2016eyeriss}
\bibfield{author}{\bibinfo{person}{{Chen, Yu-Hsin and Krishna, Tushar and Emer,
  Joel and Sze, Vivienne}}.} \bibinfo{year}{{2016}}\natexlab{}.
\newblock \showarticletitle{Eyeriss: An Energy-Efficient Reconfigurable
  Accelerator for Deep Convolutional Neural Networks}. In
  \bibinfo{booktitle}{\emph{{IEEE International Solid-State Circuits
  Conference, ISSCC 2016, Digest of Technical Papers}}}.
  \bibinfo{pages}{{262--263}}.
\newblock


\bibitem[\protect\citeauthoryear{Choquette and Gandhi}{Choquette and
  Gandhi}{2020}]%
        {choquette2020nvidia}
\bibfield{author}{\bibinfo{person}{Jack Choquette} {and} \bibinfo{person}{Wish
  Gandhi}.} \bibinfo{year}{2020}\natexlab{}.
\newblock \showarticletitle{{NVIDIA A100 GPU:} Performance \& Innovation for
  {GPU} Computing}. In \bibinfo{booktitle}{\emph{2020 IEEE Hot Chips 32
  Symposium (HCS)}}. IEEE Computer Society, \bibinfo{pages}{1--43}.
\newblock


\bibitem[\protect\citeauthoryear{Chou, Kj{\o}lstad, and Amarasinghe}{Chou
  et~al\mbox{.}}{2018}]%
        {chou2018format}
\bibfield{author}{\bibinfo{person}{Stephen Chou}, \bibinfo{person}{Fredrik
  Kj{\o}lstad}, {and} \bibinfo{person}{Saman Amarasinghe}.}
  \bibinfo{year}{2018}\natexlab{}.
\newblock \showarticletitle{Format Abstraction for Sparse Tensor Algebra
  Compilers}.
\newblock \bibinfo{journal}{\emph{Proceedings of the ACM on Programming
  Languages}} \bibinfo{volume}{2}, \bibinfo{number}{OOPSLA}
  (\bibinfo{year}{2018}), \bibinfo{pages}{1--30}.
\newblock


\bibitem[\protect\citeauthoryear{Chung, Hoe, and Mai}{Chung
  et~al\mbox{.}}{2011}]%
        {chung2011coram}
\bibfield{author}{\bibinfo{person}{Eric~S. Chung}, \bibinfo{person}{James~C.
  Hoe}, {and} \bibinfo{person}{Ken Mai}.} \bibinfo{year}{2011}\natexlab{}.
\newblock \showarticletitle{{CoRAM}: An In-Fabric Memory Architecture for
  {FPGA}-Based Computing}. In \bibinfo{booktitle}{\emph{Proceedings of the 19th
  ACM/SIGDA International Symposium on Field Programmable Gate Arrays}}
  (Monterey, CA, USA) \emph{(\bibinfo{series}{FPGA ’11})}.
  \bibinfo{publisher}{Association for Computing Machinery},
  \bibinfo{address}{New York, NY, USA}, \bibinfo{pages}{97–106}.
\newblock
\showISBNx{9781450305549}
\urldef\tempurl%
\url{https://doi.org/10.1145/1950413.1950435}
\showDOI{\tempurl}


\bibitem[\protect\citeauthoryear{Dadu, Liu, and Nowatzki}{Dadu
  et~al\mbox{.}}{2021}]%
        {dadu2021polygraph}
\bibfield{author}{\bibinfo{person}{Vidushi Dadu}, \bibinfo{person}{Sihao Liu},
  {and} \bibinfo{person}{Tony Nowatzki}.} \bibinfo{year}{2021}\natexlab{}.
\newblock \showarticletitle{PolyGraph: Exposing the Value of Flexibility for
  Graph Processing Accelerators}. In \bibinfo{booktitle}{\emph{2021 ACM/IEEE
  48th Annual International Symposium on Computer Architecture (ISCA)}}. IEEE,
  \bibinfo{pages}{595--608}.
\newblock


\bibitem[\protect\citeauthoryear{Dadu, Weng, Liu, and Nowatzki}{Dadu
  et~al\mbox{.}}{2019}]%
        {dadu2019towards}
\bibfield{author}{\bibinfo{person}{Vidushi Dadu}, \bibinfo{person}{Jian Weng},
  \bibinfo{person}{Sihao Liu}, {and} \bibinfo{person}{Tony Nowatzki}.}
  \bibinfo{year}{2019}\natexlab{}.
\newblock \showarticletitle{Towards General Purpose Acceleration by Exploiting
  Common Data-Dependence Forms}. In \bibinfo{booktitle}{\emph{Proceedings of
  the 52nd Annual IEEE/ACM International Symposium on Microarchitecture}}. ACM,
  \bibinfo{pages}{924--939}.
\newblock


\bibitem[\protect\citeauthoryear{Dai, Chi, Wang, and Yang}{Dai
  et~al\mbox{.}}{2016}]%
        {dai2016fpgp}
\bibfield{author}{\bibinfo{person}{Guohao Dai}, \bibinfo{person}{Yuze Chi},
  \bibinfo{person}{Yu Wang}, {and} \bibinfo{person}{Huazhong Yang}.}
  \bibinfo{year}{2016}\natexlab{}.
\newblock \showarticletitle{{FPGP}: Graph Processing Framework on {FPGA} a Case
  Study of Breadth-First Search}. In \bibinfo{booktitle}{\emph{Proceedings of
  the 2016 ACM/SIGDA International Symposium on Field-Programmable Gate
  Arrays}}. ACM, \bibinfo{pages}{105--110}.
\newblock


\bibitem[\protect\citeauthoryear{Dally and Towles}{Dally and Towles}{2004}]%
        {dally2004principles}
\bibfield{author}{\bibinfo{person}{William~James Dally} {and}
  \bibinfo{person}{Brian~Patrick Towles}.} \bibinfo{year}{2004}\natexlab{}.
\newblock \bibinfo{booktitle}{\emph{Principles and Practices of Interconnection
  Networks}}.
\newblock \bibinfo{publisher}{Elsevier}.
\newblock


\bibitem[\protect\citeauthoryear{Dave, Baghdadi, Nowatzki, Avancha,
  Shrivastava, and Li}{Dave et~al\mbox{.}}{2020}]%
        {dave2020hardware}
\bibfield{author}{\bibinfo{person}{Shail Dave}, \bibinfo{person}{Riyadh
  Baghdadi}, \bibinfo{person}{Tony Nowatzki}, \bibinfo{person}{Sasikanth
  Avancha}, \bibinfo{person}{Aviral Shrivastava}, {and} \bibinfo{person}{Baoxin
  Li}.} \bibinfo{year}{2020}\natexlab{}.
\newblock \showarticletitle{Hardware Acceleration of Sparse and Irregular
  Tensor Computations of ML Models: A Survey and Insights}.
\newblock \bibinfo{journal}{\emph{arXiv preprint arXiv:2007.00864}}
  (\bibinfo{year}{2020}).
\newblock


\bibitem[\protect\citeauthoryear{Davis and Hu}{Davis and Hu}{2011}]%
        {davis2011university}
\bibfield{author}{\bibinfo{person}{Timothy~A Davis} {and}
  \bibinfo{person}{Yifan Hu}.} \bibinfo{year}{2011}\natexlab{}.
\newblock \showarticletitle{The {University of Florida} Sparse Matrix
  Collection}.
\newblock \bibinfo{journal}{\emph{ACM Transactions on Mathematical Software
  (TOMS)}} \bibinfo{volume}{38}, \bibinfo{number}{1} (\bibinfo{year}{2011}),
  \bibinfo{pages}{1}.
\newblock


\bibitem[\protect\citeauthoryear{Goldstein, Schmit, Moe, Budiu, Cadambi,
  Taylor, and Laufer}{Goldstein et~al\mbox{.}}{1999}]%
        {goldstein1999piperench}
\bibfield{author}{\bibinfo{person}{Seth~Copen Goldstein},
  \bibinfo{person}{Herman Schmit}, \bibinfo{person}{Matthew Moe},
  \bibinfo{person}{Mihai Budiu}, \bibinfo{person}{Srihari Cadambi},
  \bibinfo{person}{R~Reed Taylor}, {and} \bibinfo{person}{Ronald Laufer}.}
  \bibinfo{year}{1999}\natexlab{}.
\newblock \showarticletitle{{PipeRench}: A Coprocessor for Streaming Multimedia
  Acceleration}. In \bibinfo{booktitle}{\emph{Proceedings of the 26th
  International Symposium on Computer Architecture (Cat. No. 99CB36367)}}.
  IEEE, \bibinfo{pages}{28--39}.
\newblock


\bibitem[\protect\citeauthoryear{Govindaraju, Ho, Nowatzki, Chhugani, Satish,
  Sankaralingam, and Kim}{Govindaraju et~al\mbox{.}}{2012}]%
        {govindaraju2012dyser}
\bibfield{author}{\bibinfo{person}{Venkatraman Govindaraju},
  \bibinfo{person}{Chen-Han Ho}, \bibinfo{person}{Tony Nowatzki},
  \bibinfo{person}{Jatin Chhugani}, \bibinfo{person}{Nadathur Satish},
  \bibinfo{person}{Karthikeyan Sankaralingam}, {and} \bibinfo{person}{Changkyu
  Kim}.} \bibinfo{year}{2012}\natexlab{}.
\newblock \showarticletitle{DySER: Unifying Functionality and Parallelism
  Specialization for Energy-Efficient Computing}.
\newblock \bibinfo{journal}{\emph{IEEE Micro}} \bibinfo{volume}{32},
  \bibinfo{number}{5} (\bibinfo{year}{2012}), \bibinfo{pages}{38--51}.
\newblock


\bibitem[\protect\citeauthoryear{Gustavson}{Gustavson}{1978}]%
        {10.1145/355791.355796}
\bibfield{author}{\bibinfo{person}{Fred~G. Gustavson}.}
  \bibinfo{year}{1978}\natexlab{}.
\newblock \showarticletitle{Two Fast Algorithms for Sparse Matrices:
  Multiplication and Permuted Transposition}.
\newblock \bibinfo{journal}{\emph{ACM Trans. Math. Softw.}}
  \bibinfo{volume}{4}, \bibinfo{number}{3} (\bibinfo{date}{Sept.}
  \bibinfo{year}{1978}), \bibinfo{pages}{250–269}.
\newblock
\showISSN{0098-3500}
\urldef\tempurl%
\url{https://doi.org/10.1145/355791.355796}
\showDOI{\tempurl}


\bibitem[\protect\citeauthoryear{Ham, Wu, Sundaram, Satish, and Martonosi}{Ham
  et~al\mbox{.}}{2016}]%
        {ham2016graphicionado}
\bibfield{author}{\bibinfo{person}{Tae~Jun Ham}, \bibinfo{person}{Lisa Wu},
  \bibinfo{person}{Narayanan Sundaram}, \bibinfo{person}{Nadathur Satish},
  {and} \bibinfo{person}{Margaret Martonosi}.} \bibinfo{year}{2016}\natexlab{}.
\newblock \showarticletitle{Graphicionado: A High-Performance and
  Energy-Efficient Accelerator for Graph Analytics}. In
  \bibinfo{booktitle}{\emph{2016 49th Annual IEEE/ACM International Symposium
  on Microarchitecture (MICRO)}}. IEEE, \bibinfo{pages}{1--13}.
\newblock


\bibitem[\protect\citeauthoryear{Han, Liu, Mao, Pu, Pedram, Horowitz, and
  Dally}{Han et~al\mbox{.}}{2016}]%
        {han2016eie}
\bibfield{author}{\bibinfo{person}{Song Han}, \bibinfo{person}{Xingyu Liu},
  \bibinfo{person}{Huizi Mao}, \bibinfo{person}{Jing Pu},
  \bibinfo{person}{Ardavan Pedram}, \bibinfo{person}{Mark~A Horowitz}, {and}
  \bibinfo{person}{William~J Dally}.} \bibinfo{year}{2016}\natexlab{}.
\newblock \showarticletitle{{EIE}: Efficient Inference Engine on Compressed
  Deep Neural Network}. In \bibinfo{booktitle}{\emph{2016 ACM/IEEE 43rd Annual
  International Symposium on Computer Architecture (ISCA)}}. IEEE,
  \bibinfo{pages}{243--254}.
\newblock


\bibitem[\protect\citeauthoryear{Han, Mao, and Dally}{Han
  et~al\mbox{.}}{2015}]%
        {han2015deep}
\bibfield{author}{\bibinfo{person}{Song Han}, \bibinfo{person}{Huizi Mao},
  {and} \bibinfo{person}{William~J Dally}.} \bibinfo{year}{2015}\natexlab{}.
\newblock \showarticletitle{Deep compression: Compressing Deep Neural Networks
  with Pruning, Trained Quantization and Huffman Coding}.
\newblock \bibinfo{journal}{\emph{arXiv preprint arXiv:1510.00149}}
  (\bibinfo{year}{2015}).
\newblock


\bibitem[\protect\citeauthoryear{He, Zhang, Ren, and Sun}{He
  et~al\mbox{.}}{2016}]%
        {he2016deep}
\bibfield{author}{\bibinfo{person}{Kaiming He}, \bibinfo{person}{Xiangyu
  Zhang}, \bibinfo{person}{Shaoqing Ren}, {and} \bibinfo{person}{Jian Sun}.}
  \bibinfo{year}{2016}\natexlab{}.
\newblock \showarticletitle{Deep Residual Learning for Image Recognition}. In
  \bibinfo{booktitle}{\emph{Proceedings of the IEEE conference on computer
  vision and pattern recognition}}. \bibinfo{pages}{770--778}.
\newblock


\bibitem[\protect\citeauthoryear{{He Qi}, {Ayorinde}, {Yu Huang}, and
  {Calhoun}}{{He Qi} et~al\mbox{.}}{2015}]%
        {lowswing}
\bibfield{author}{\bibinfo{person}{{He Qi}}, \bibinfo{person}{O. {Ayorinde}},
  \bibinfo{person}{{Yu Huang}}, {and} \bibinfo{person}{B. {Calhoun}}.}
  \bibinfo{year}{2015}\natexlab{}.
\newblock \showarticletitle{Optimizing Energy Efficient Low-Swing Interconnect
  for Sub-Threshold {FPGAs}}. In \bibinfo{booktitle}{\emph{2015 25th
  International Conference on Field Programmable Logic and Applications
  (FPL)}}. \bibinfo{pages}{1--4}.
\newblock
\urldef\tempurl%
\url{https://doi.org/10.1109/FPL.2015.7293979}
\showDOI{\tempurl}


\bibitem[\protect\citeauthoryear{Heckbert}{Heckbert}{1982}]%
        {heckbert1982}
\bibfield{author}{\bibinfo{person}{Paul Heckbert}.}
  \bibinfo{year}{1982}\natexlab{}.
\newblock \showarticletitle{Color Image Quantization for Frame Buffer Display}.
\newblock \bibinfo{journal}{\emph{SIGGRAPH Comput. Graph.}}
  \bibinfo{volume}{16}, \bibinfo{number}{3} (\bibinfo{date}{July}
  \bibinfo{year}{1982}), \bibinfo{pages}{297–307}.
\newblock
\showISSN{0097-8930}
\urldef\tempurl%
\url{https://doi.org/10.1145/965145.801294}
\showDOI{\tempurl}


\bibitem[\protect\citeauthoryear{Hegde, Asghari-Moghaddam, Pellauer, Crago,
  Jaleel, Solomonik, Emer, and Fletcher}{Hegde et~al\mbox{.}}{2019}]%
        {hegde2019extensor}
\bibfield{author}{\bibinfo{person}{Kartik Hegde}, \bibinfo{person}{Hadi
  Asghari-Moghaddam}, \bibinfo{person}{Michael Pellauer}, \bibinfo{person}{Neal
  Crago}, \bibinfo{person}{Aamer Jaleel}, \bibinfo{person}{Edgar Solomonik},
  \bibinfo{person}{Joel Emer}, {and} \bibinfo{person}{Christopher~W.
  Fletcher}.} \bibinfo{year}{2019}\natexlab{}.
\newblock \showarticletitle{{ExTensor}: An Accelerator for Sparse Tensor
  Algebra}. In \bibinfo{booktitle}{\emph{Proceedings of the 52nd Annual
  IEEE/ACM International Symposium on Microarchitecture}} (Columbus, OH, USA)
  \emph{(\bibinfo{series}{MICRO ’52})}. \bibinfo{publisher}{Association for
  Computing Machinery}, \bibinfo{address}{New York, NY, USA},
  \bibinfo{pages}{319–333}.
\newblock
\showISBNx{9781450369381}
\urldef\tempurl%
\url{https://doi.org/10.1145/3352460.3358275}
\showDOI{\tempurl}


\bibitem[\protect\citeauthoryear{Iandola, Han, Moskewicz, Ashraf, Dally, and
  Keutzer}{Iandola et~al\mbox{.}}{2016}]%
        {iandola2016squeezenet}
\bibfield{author}{\bibinfo{person}{Forrest~N Iandola}, \bibinfo{person}{Song
  Han}, \bibinfo{person}{Matthew~W Moskewicz}, \bibinfo{person}{Khalid Ashraf},
  \bibinfo{person}{William~J Dally}, {and} \bibinfo{person}{Kurt Keutzer}.}
  \bibinfo{year}{2016}\natexlab{}.
\newblock \showarticletitle{{SqueezeNet}: {AlexNet}-Level Accuracy with 50x
  Fewer Parameters and< 0.5 {MB} Model Size}.
\newblock \bibinfo{journal}{\emph{arXiv preprint arXiv:1602.07360}}
  (\bibinfo{year}{2016}).
\newblock


\bibitem[\protect\citeauthoryear{Intel}{Intel}{[n.d.]a}]%
        {mkl}
\bibfield{author}{\bibinfo{person}{Intel}.} \bibinfo{year}{[n.d.]}\natexlab{a}.
\newblock \bibinfo{booktitle}{\emph{Intel Math Kernel Library}}.
\newblock
\urldef\tempurl%
\url{https://software.intel.com/en-us/mkl}
\showURL{%
\tempurl}


\bibitem[\protect\citeauthoryear{Intel}{Intel}{[n.d.]b}]%
        {xeon8890v3}
\bibfield{author}{\bibinfo{person}{Intel}.} \bibinfo{year}{[n.d.]}\natexlab{b}.
\newblock \bibinfo{booktitle}{\emph{Intel {Xeon} Processor {E7-8890} v3}}.
\newblock
\urldef\tempurl%
\url{https://ark.intel.com/content/www/us/en/ark/products/84685/intel-xeon-processor-e7-8890-v3-45m-cache-2-50-ghz.html}
\showURL{%
\tempurl}


\bibitem[\protect\citeauthoryear{Jiang, Becker, Michelogiannakis, and
  Dally}{Jiang et~al\mbox{.}}{2011}]%
        {jiang2011performance}
\bibfield{author}{\bibinfo{person}{Nan Jiang}, \bibinfo{person}{Daniel Becker},
  \bibinfo{person}{George Michelogiannakis}, {and} \bibinfo{person}{William~J.
  Dally}.} \bibinfo{year}{2011}\natexlab{}.
\newblock \showarticletitle{Performance Implications of Age-Based Allocation in
  On-Chip-Networks}.
\newblock  (\bibinfo{year}{2011}).
\newblock


\bibitem[\protect\citeauthoryear{Jouppi, Young, Patil, Patterson, Agrawal,
  Bajwa, Bates, Bhatia, Boden, Borchers, et~al\mbox{.}}{Jouppi
  et~al\mbox{.}}{2017}]%
        {jouppi2017datacenter}
\bibfield{author}{\bibinfo{person}{Norman~P Jouppi}, \bibinfo{person}{Cliff
  Young}, \bibinfo{person}{Nishant Patil}, \bibinfo{person}{David Patterson},
  \bibinfo{person}{Gaurav Agrawal}, \bibinfo{person}{Raminder Bajwa},
  \bibinfo{person}{Sarah Bates}, \bibinfo{person}{Suresh Bhatia},
  \bibinfo{person}{Nan Boden}, \bibinfo{person}{Al Borchers}, {et~al\mbox{.}}}
  \bibinfo{year}{2017}\natexlab{}.
\newblock \showarticletitle{In-Datacenter Performance Analysis of a Tensor
  Processing Unit}. In \bibinfo{booktitle}{\emph{Proceedings of the 44th Annual
  International Symposium on Computer Architecture}}. \bibinfo{pages}{1--12}.
\newblock


\bibitem[\protect\citeauthoryear{Kanellopoulos, Vijaykumar, Giannoula, Azizi,
  Koppula, Ghiasi, Shahroodi, Luna, and Mutlu}{Kanellopoulos
  et~al\mbox{.}}{2019}]%
        {kanellopoulos2019}
\bibfield{author}{\bibinfo{person}{Konstantinos Kanellopoulos},
  \bibinfo{person}{Nandita Vijaykumar}, \bibinfo{person}{Christina Giannoula},
  \bibinfo{person}{Roknoddin Azizi}, \bibinfo{person}{Skanda Koppula},
  \bibinfo{person}{Nika~Mansouri Ghiasi}, \bibinfo{person}{Taha Shahroodi},
  \bibinfo{person}{Juan~Gomez Luna}, {and} \bibinfo{person}{Onur Mutlu}.}
  \bibinfo{year}{2019}\natexlab{}.
\newblock \showarticletitle{SMASH: Co-Designing Software Compression and
  Hardware-Accelerated Indexing for Efficient Sparse Matrix Operations}. In
  \bibinfo{booktitle}{\emph{Proceedings of the 52nd Annual IEEE/ACM
  International Symposium on Microarchitecture}} (Columbus, OH, USA)
  \emph{(\bibinfo{series}{MICRO ’52})}. \bibinfo{publisher}{Association for
  Computing Machinery}, \bibinfo{address}{New York, NY, USA},
  \bibinfo{pages}{600–614}.
\newblock
\showISBNx{9781450369381}
\urldef\tempurl%
\url{https://doi.org/10.1145/3352460.3358286}
\showDOI{\tempurl}


\bibitem[\protect\citeauthoryear{Karypis and Kumar}{Karypis and Kumar}{1998}]%
        {karypis1998fast}
\bibfield{author}{\bibinfo{person}{George Karypis} {and} \bibinfo{person}{Vipin
  Kumar}.} \bibinfo{year}{1998}\natexlab{}.
\newblock \showarticletitle{A Fast and High Quality Multilevel Scheme for
  Partitioning Irregular Graphs}.
\newblock \bibinfo{journal}{\emph{SIAM Journal on scientific Computing}}
  \bibinfo{volume}{20}, \bibinfo{number}{1} (\bibinfo{year}{1998}),
  \bibinfo{pages}{359--392}.
\newblock


\bibitem[\protect\citeauthoryear{Kepner, Aaltonen, Bader, Bulu{\c{c}},
  Franchetti, Gilbert, Hutchison, Kumar, Lumsdaine, Meyerhenke,
  et~al\mbox{.}}{Kepner et~al\mbox{.}}{2016}]%
        {kepner2016mathematical}
\bibfield{author}{\bibinfo{person}{Jeremy Kepner}, \bibinfo{person}{Peter
  Aaltonen}, \bibinfo{person}{David Bader}, \bibinfo{person}{Aydin
  Bulu{\c{c}}}, \bibinfo{person}{Franz Franchetti}, \bibinfo{person}{John
  Gilbert}, \bibinfo{person}{Dylan Hutchison}, \bibinfo{person}{Manoj Kumar},
  \bibinfo{person}{Andrew Lumsdaine}, \bibinfo{person}{Henning Meyerhenke},
  {et~al\mbox{.}}} \bibinfo{year}{2016}\natexlab{}.
\newblock \showarticletitle{Mathematical Foundations of the {GraphBLAS}}. In
  \bibinfo{booktitle}{\emph{2016 IEEE High Performance Extreme Computing
  Conference (HPEC)}}. IEEE, \bibinfo{pages}{1--9}.
\newblock


\bibitem[\protect\citeauthoryear{Keskar, Mudigere, Nocedal, Smelyanskiy, and
  Tang}{Keskar et~al\mbox{.}}{2017}]%
        {keskar2017largebatch}
\bibfield{author}{\bibinfo{person}{Nitish~Shirish Keskar},
  \bibinfo{person}{Dheevatsa Mudigere}, \bibinfo{person}{Jorge Nocedal},
  \bibinfo{person}{Mikhail Smelyanskiy}, {and} \bibinfo{person}{Ping Tak~Peter
  Tang}.} \bibinfo{year}{2017}\natexlab{}.
\newblock \bibinfo{title}{On Large-Batch Training for Deep Learning:
  Generalization Gap and Sharp Minima}.
\newblock
\newblock
\showeprint[arxiv]{1609.04836}~[cs.LG]


\bibitem[\protect\citeauthoryear{Kim, Yang, and Mutlu}{Kim
  et~al\mbox{.}}{2015}]%
        {kim2015ramulator}
\bibfield{author}{\bibinfo{person}{Yoongu Kim}, \bibinfo{person}{Weikun Yang},
  {and} \bibinfo{person}{Onur Mutlu}.} \bibinfo{year}{2015}\natexlab{}.
\newblock \showarticletitle{Ramulator: A Fast and Extensible {DRAM} Simulator}.
\newblock \bibinfo{journal}{\emph{IEEE Computer Architecture Letters}}
  \bibinfo{volume}{15}, \bibinfo{number}{1} (\bibinfo{year}{2015}),
  \bibinfo{pages}{45--49}.
\newblock


\bibitem[\protect\citeauthoryear{Kj{\o}lstad, Chou, Lugato, Kamil, and
  Amarasinghe}{Kj{\o}lstad et~al\mbox{.}}{2017}]%
        {taco}
\bibfield{author}{\bibinfo{person}{Fredrik Kj{\o}lstad},
  \bibinfo{person}{Stephen Chou}, \bibinfo{person}{David Lugato},
  \bibinfo{person}{Shoaib Kamil}, {and} \bibinfo{person}{Saman Amarasinghe}.}
  \bibinfo{year}{2017}\natexlab{}.
\newblock \showarticletitle{{TACO}: A Tool to Generate Tensor Algebra Kernels}.
  In \bibinfo{booktitle}{\emph{2017 32nd IEEE/ACM International Conference on
  Automated Software Engineering (ASE)}}. IEEE, \bibinfo{pages}{943--948}.
\newblock


\bibitem[\protect\citeauthoryear{Kj{\o}lstad}{Kj{\o}lstad}{2020}]%
        {kjolstad2020sparse}
\bibfield{author}{\bibinfo{person}{Fredrik~Berg Kj{\o}lstad}.}
  \bibinfo{year}{2020}\natexlab{}.
\newblock \emph{\bibinfo{title}{Sparse Tensor Algebra Compilation}}.
\newblock \bibinfo{thesistype}{Ph.D. Dissertation}.
  \bibinfo{school}{Massachusetts Institute of Technology}.
\newblock


\bibitem[\protect\citeauthoryear{Koeplinger, Feldman, Prabhakar, Zhang, Hadjis,
  Fiszel, Zhao, Nardi, Pedram, Kozyrakis, et~al\mbox{.}}{Koeplinger
  et~al\mbox{.}}{2018}]%
        {koeplinger2018spatial}
\bibfield{author}{\bibinfo{person}{David Koeplinger}, \bibinfo{person}{Matthew
  Feldman}, \bibinfo{person}{Raghu Prabhakar}, \bibinfo{person}{Yaqi Zhang},
  \bibinfo{person}{Stefan Hadjis}, \bibinfo{person}{Ruben Fiszel},
  \bibinfo{person}{Tian Zhao}, \bibinfo{person}{Luigi Nardi},
  \bibinfo{person}{Ardavan Pedram}, \bibinfo{person}{Christos Kozyrakis},
  {et~al\mbox{.}}} \bibinfo{year}{2018}\natexlab{}.
\newblock \showarticletitle{Spatial: A Language and Compiler for Application
  Accelerators}. In \bibinfo{booktitle}{\emph{ACM SIGPLAN Notices}},
  Vol.~\bibinfo{volume}{53}. ACM, \bibinfo{pages}{296--311}.
\newblock


\bibitem[\protect\citeauthoryear{Komuravelli, Sinclair, Alsop, Huzaifa,
  Kotsifakou, Srivastava, Adve, and Adve}{Komuravelli et~al\mbox{.}}{2015}]%
        {komuravelli2015stash}
\bibfield{author}{\bibinfo{person}{Rakesh Komuravelli},
  \bibinfo{person}{Matthew~D. Sinclair}, \bibinfo{person}{Johnathan Alsop},
  \bibinfo{person}{Muhammad Huzaifa}, \bibinfo{person}{Maria Kotsifakou},
  \bibinfo{person}{Prakalp Srivastava}, \bibinfo{person}{Sarita~V. Adve}, {and}
  \bibinfo{person}{Vikram~S. Adve}.} \bibinfo{year}{2015}\natexlab{}.
\newblock \showarticletitle{Stash: Have Your Scratchpad and Cache It Too}. In
  \bibinfo{booktitle}{\emph{Proceedings of the 42nd Annual International
  Symposium on Computer Architecture}} (Portland, Oregon)
  \emph{(\bibinfo{series}{ISCA ’15})}. \bibinfo{publisher}{Association for
  Computing Machinery}, \bibinfo{address}{New York, NY, USA},
  \bibinfo{pages}{707–719}.
\newblock
\showISBNx{9781450334020}
\urldef\tempurl%
\url{https://doi.org/10.1145/2749469.2750374}
\showDOI{\tempurl}


\bibitem[\protect\citeauthoryear{Krajcevski, Pratapa, and Manocha}{Krajcevski
  et~al\mbox{.}}{2016}]%
        {krajcevski2016}
\bibfield{author}{\bibinfo{person}{Pavel Krajcevski}, \bibinfo{person}{Srihari
  Pratapa}, {and} \bibinfo{person}{Dinesh Manocha}.}
  \bibinfo{year}{2016}\natexlab{}.
\newblock \showarticletitle{{GST}: {GPU}-Decodable Supercompressed Textures}.
\newblock \bibinfo{journal}{\emph{ACM Trans. Graph.}} \bibinfo{volume}{35},
  \bibinfo{number}{6}, Article \bibinfo{articleno}{230} (\bibinfo{date}{Nov.}
  \bibinfo{year}{2016}), \bibinfo{numpages}{10}~pages.
\newblock
\showISSN{0730-0301}
\urldef\tempurl%
\url{https://doi.org/10.1145/2980179.2982439}
\showDOI{\tempurl}


\bibitem[\protect\citeauthoryear{Leskovec, Lang, Dasgupta, and
  Mahoney}{Leskovec et~al\mbox{.}}{2009}]%
        {leskovec2009community}
\bibfield{author}{\bibinfo{person}{Jure Leskovec}, \bibinfo{person}{Kevin~J
  Lang}, \bibinfo{person}{Anirban Dasgupta}, {and} \bibinfo{person}{Michael~W
  Mahoney}.} \bibinfo{year}{2009}\natexlab{}.
\newblock \showarticletitle{Community Structure in Large Networks: Natural
  Cluster Sizes and the Absence of Large Well-Defined Clusters}.
\newblock \bibinfo{journal}{\emph{Internet Mathematics}} \bibinfo{volume}{6},
  \bibinfo{number}{1} (\bibinfo{year}{2009}), \bibinfo{pages}{29--123}.
\newblock


\bibitem[\protect\citeauthoryear{Martins, Matos, Ribas, Reis, Schlinker, Rech,
  and Michelsen}{Martins et~al\mbox{.}}{2015}]%
        {freepdk15}
\bibfield{author}{\bibinfo{person}{Mayler Martins}, \bibinfo{person}{Jody~Maick
  Matos}, \bibinfo{person}{Renato~P. Ribas}, \bibinfo{person}{Andr\'{e} Reis},
  \bibinfo{person}{Guilherme Schlinker}, \bibinfo{person}{Lucio Rech}, {and}
  \bibinfo{person}{Jens Michelsen}.} \bibinfo{year}{2015}\natexlab{}.
\newblock \showarticletitle{Open Cell Library in 15nm {FreePDK} Technology}. In
  \bibinfo{booktitle}{\emph{Proceedings of the 2015 Symposium on International
  Symposium on Physical Design}} (Monterey, California, USA)
  \emph{(\bibinfo{series}{ISPD '15})}. \bibinfo{publisher}{Association for
  Computing Machinery}, \bibinfo{address}{New York, NY, USA},
  \bibinfo{pages}{171--178}.
\newblock
\showISBNx{9781450333993}
\urldef\tempurl%
\url{https://doi.org/10.1145/2717764.2717783}
\showDOI{\tempurl}


\bibitem[\protect\citeauthoryear{McGettrick, Geraghty, and McElroy}{McGettrick
  et~al\mbox{.}}{2008}]%
        {mcgettrick2008fpga}
\bibfield{author}{\bibinfo{person}{Seamas McGettrick}, \bibinfo{person}{Dermot
  Geraghty}, {and} \bibinfo{person}{Ciaran McElroy}.}
  \bibinfo{year}{2008}\natexlab{}.
\newblock \showarticletitle{An {FPGA} Architecture for the {PageRank}
  Eigenvector Problem}. In \bibinfo{booktitle}{\emph{2008 International
  Conference on Field Programmable Logic and Applications}}. IEEE,
  \bibinfo{pages}{523--526}.
\newblock


\bibitem[\protect\citeauthoryear{McKeown}{McKeown}{1999}]%
        {mckeown1999islip}
\bibfield{author}{\bibinfo{person}{Nick McKeown}.}
  \bibinfo{year}{1999}\natexlab{}.
\newblock \showarticletitle{The {iSLIP} Scheduling Algorithm for Input-Queued
  Switches}.
\newblock \bibinfo{journal}{\emph{IEEE/ACM transactions on networking}}
  \bibinfo{volume}{7}, \bibinfo{number}{2} (\bibinfo{year}{1999}),
  \bibinfo{pages}{188--201}.
\newblock


\bibitem[\protect\citeauthoryear{Nurvitadhi, Weisz, Wang, Hurkat, Nguyen, Hoe,
  Mart{\'\i}nez, and Guestrin}{Nurvitadhi et~al\mbox{.}}{2014}]%
        {nurvitadhi2014graphgen}
\bibfield{author}{\bibinfo{person}{Eriko Nurvitadhi}, \bibinfo{person}{Gabriel
  Weisz}, \bibinfo{person}{Yu Wang}, \bibinfo{person}{Skand Hurkat},
  \bibinfo{person}{Marie Nguyen}, \bibinfo{person}{James~C Hoe},
  \bibinfo{person}{Jos{\'e}~F Mart{\'\i}nez}, {and} \bibinfo{person}{Carlos
  Guestrin}.} \bibinfo{year}{2014}\natexlab{}.
\newblock \showarticletitle{{GraphGen}: An {FPGA} Framework for Vertex-Centric
  Graph Computation}. In \bibinfo{booktitle}{\emph{2014 IEEE 22nd Annual
  International Symposium on Field-Programmable Custom Computing Machines}}.
  IEEE, \bibinfo{pages}{25--28}.
\newblock


\bibitem[\protect\citeauthoryear{Nvidia}{Nvidia}{[n.d.]}]%
        {v100datasheet}
\bibfield{author}{\bibinfo{person}{Nvidia}.} \bibinfo{year}{[n.d.]}\natexlab{}.
\newblock \bibinfo{booktitle}{\emph{Nvidia {Tesla V100 GPU} Architecture}}.
\newblock
\urldef\tempurl%
\url{https://images.nvidia.com/content/volta-architecture/pdf/volta-architecture-whitepaper.pdf}
\showURL{%
\tempurl}


\bibitem[\protect\citeauthoryear{Nvidia}{Nvidia}{2019}]%
        {cuSparse}
\bibfield{author}{\bibinfo{person}{Nvidia}.} \bibinfo{year}{2019}\natexlab{}.
\newblock \bibinfo{booktitle}{\emph{The API reference guide for cuSPARSE}}.
\newblock


\bibitem[\protect\citeauthoryear{Oguntebi and Olukotun}{Oguntebi and
  Olukotun}{2016}]%
        {oguntebi2016graphops}
\bibfield{author}{\bibinfo{person}{Tayo Oguntebi} {and} \bibinfo{person}{Kunle
  Olukotun}.} \bibinfo{year}{2016}\natexlab{}.
\newblock \showarticletitle{{GraphOps}: A Dataflow Library for Graph Analytics
  Acceleration}. In \bibinfo{booktitle}{\emph{Proceedings of the 2016 ACM/SIGDA
  International Symposium on Field-Programmable Gate Arrays}}. ACM,
  \bibinfo{pages}{111--117}.
\newblock


\bibitem[\protect\citeauthoryear{Pal, Beaumont, Park, Amarnath, Feng,
  Chakrabarti, Kim, Blaauw, Mudge, and Dreslinski}{Pal et~al\mbox{.}}{2018}]%
        {pal2018outerspace}
\bibfield{author}{\bibinfo{person}{Subhankar Pal}, \bibinfo{person}{Jonathan
  Beaumont}, \bibinfo{person}{Dong-Hyeon Park}, \bibinfo{person}{Aporva
  Amarnath}, \bibinfo{person}{Siying Feng}, \bibinfo{person}{Chaitali
  Chakrabarti}, \bibinfo{person}{Hun-Seok Kim}, \bibinfo{person}{David Blaauw},
  \bibinfo{person}{Trevor Mudge}, {and} \bibinfo{person}{Ronald Dreslinski}.}
  \bibinfo{year}{2018}\natexlab{}.
\newblock \showarticletitle{OuterSPACE: An Outer Product Based Sparse Matrix
  Multiplication Accelerator}. In \bibinfo{booktitle}{\emph{2018 IEEE
  International Symposium on High Performance Computer Architecture (HPCA)}}.
  IEEE, \bibinfo{pages}{724--736}.
\newblock


\bibitem[\protect\citeauthoryear{Parashar, Rhu, Mukkara, Puglielli, Venkatesan,
  Khailany, Emer, Keckler, and Dally}{Parashar et~al\mbox{.}}{2017}]%
        {parashar2017scnn}
\bibfield{author}{\bibinfo{person}{Angshuman Parashar}, \bibinfo{person}{Minsoo
  Rhu}, \bibinfo{person}{Anurag Mukkara}, \bibinfo{person}{Antonio Puglielli},
  \bibinfo{person}{Rangharajan Venkatesan}, \bibinfo{person}{Brucek Khailany},
  \bibinfo{person}{Joel Emer}, \bibinfo{person}{Stephen~W Keckler}, {and}
  \bibinfo{person}{William~J Dally}.} \bibinfo{year}{2017}\natexlab{}.
\newblock \showarticletitle{{SCNN}: An Accelerator for Compressed-Sparse
  Convolutional Neural Networks}. In \bibinfo{booktitle}{\emph{2017 ACM/IEEE
  44th Annual International Symposium on Computer Architecture (ISCA)}}. IEEE,
  \bibinfo{pages}{27--40}.
\newblock


\bibitem[\protect\citeauthoryear{Pellauer, Shao, Clemons, Crago, Hegde,
  Venkatesan, Keckler, Fletcher, and Emer}{Pellauer et~al\mbox{.}}{2019}]%
        {pellauer2019buffet}
\bibfield{author}{\bibinfo{person}{Michael Pellauer},
  \bibinfo{person}{Yakun~Sophia Shao}, \bibinfo{person}{Jason Clemons},
  \bibinfo{person}{Neal Crago}, \bibinfo{person}{Kartik Hegde},
  \bibinfo{person}{Rangharajan Venkatesan}, \bibinfo{person}{Stephen~W.
  Keckler}, \bibinfo{person}{Christopher~W. Fletcher}, {and}
  \bibinfo{person}{Joel Emer}.} \bibinfo{year}{2019}\natexlab{}.
\newblock \showarticletitle{Buffets: An Efficient and Composable Storage Idiom
  for Explicit Decoupled Data Orchestration}. In
  \bibinfo{booktitle}{\emph{Proceedings of the Twenty-Fourth International
  Conference on Architectural Support for Programming Languages and Operating
  Systems}} (Providence, RI, USA) \emph{(\bibinfo{series}{ASPLOS ’19})}.
  \bibinfo{publisher}{Association for Computing Machinery},
  \bibinfo{address}{New York, NY, USA}, \bibinfo{pages}{137–151}.
\newblock
\showISBNx{9781450362405}
\urldef\tempurl%
\url{https://doi.org/10.1145/3297858.3304025}
\showDOI{\tempurl}


\bibitem[\protect\citeauthoryear{Prabhakar, Zhang, Koeplinger, Feldman, Zhao,
  Hadjis, Pedram, Kozyrakis, and Olukotun}{Prabhakar et~al\mbox{.}}{2017}]%
        {prabhakar2017plasticine}
\bibfield{author}{\bibinfo{person}{Raghu Prabhakar}, \bibinfo{person}{Yaqi
  Zhang}, \bibinfo{person}{David Koeplinger}, \bibinfo{person}{Matt Feldman},
  \bibinfo{person}{Tian Zhao}, \bibinfo{person}{Stefan Hadjis},
  \bibinfo{person}{Ardavan Pedram}, \bibinfo{person}{Christos Kozyrakis}, {and}
  \bibinfo{person}{Kunle Olukotun}.} \bibinfo{year}{2017}\natexlab{}.
\newblock \showarticletitle{Plasticine: A Reconfigurable Architecture for
  Parallel Patterns}. In \bibinfo{booktitle}{\emph{2017 ACM/IEEE 44th Annual
  International Symposium on Computer Architecture (ISCA)}}. IEEE,
  \bibinfo{pages}{389--402}.
\newblock


\bibitem[\protect\citeauthoryear{Ripeanu and Foster}{Ripeanu and
  Foster}{2002}]%
        {ripeanu2002mapping}
\bibfield{author}{\bibinfo{person}{Matei Ripeanu} {and} \bibinfo{person}{Ian
  Foster}.} \bibinfo{year}{2002}\natexlab{}.
\newblock \showarticletitle{Mapping the {Gnutella} Network: Macroscopic
  Properties of Large-Scale Peer-to-Peer Systems}. In
  \bibinfo{booktitle}{\emph{international workshop on peer-to-peer systems}}.
  Springer, \bibinfo{pages}{85--93}.
\newblock


\bibitem[\protect\citeauthoryear{SambaNova}{SambaNova}{2021}]%
        {sambanova}
\bibfield{author}{\bibinfo{person}{SambaNova}.}
  \bibinfo{year}{2021}\natexlab{}.
\newblock \bibinfo{title}{Accelerated Computing with a Reconfigurable Dataflow
  Architecture}.
\newblock
\newblock
\urldef\tempurl%
\url{https://sambanova.ai/wp-content/uploads/2021/04/SambaNova\_RDA\_Whitepaper.pdf}
\showURL{%
\tempurl}


\bibitem[\protect\citeauthoryear{Seznec and Bodin}{Seznec and Bodin}{1993}]%
        {seznec1993skewed}
\bibfield{author}{\bibinfo{person}{Andr{\'e} Seznec} {and}
  \bibinfo{person}{Francois Bodin}.} \bibinfo{year}{1993}\natexlab{}.
\newblock \showarticletitle{Skewed-Associative Caches}. In
  \bibinfo{booktitle}{\emph{International Conference on Parallel Architectures
  and Languages Europe}}. Springer, \bibinfo{pages}{305--316}.
\newblock


\bibitem[\protect\citeauthoryear{Shim, Chang, and Pedram}{Shim
  et~al\mbox{.}}{2004}]%
        {shim2004}
\bibfield{author}{\bibinfo{person}{Hojun Shim}, \bibinfo{person}{Naehyuck
  Chang}, {and} \bibinfo{person}{Massoud Pedram}.}
  \bibinfo{year}{2004}\natexlab{}.
\newblock \showarticletitle{A Compressed Frame Buffer to Reduce Display Power
  Consumption in Mobile Systems}. In \bibinfo{booktitle}{\emph{Proceedings of
  the 2004 Asia and South Pacific Design Automation Conference}} (Yokohama,
  Japan) \emph{(\bibinfo{series}{ASP-DAC ’04})}. \bibinfo{publisher}{IEEE
  Press}, \bibinfo{pages}{818–823}.
\newblock
\showISBNx{0780381750}


\bibitem[\protect\citeauthoryear{Song, Zhuo, Qian, Li, and Chen}{Song
  et~al\mbox{.}}{2018}]%
        {song2018graphr}
\bibfield{author}{\bibinfo{person}{Linghao Song}, \bibinfo{person}{Youwei
  Zhuo}, \bibinfo{person}{Xuehai Qian}, \bibinfo{person}{Hai Li}, {and}
  \bibinfo{person}{Yiran Chen}.} \bibinfo{year}{2018}\natexlab{}.
\newblock \showarticletitle{{GraphR:} Accelerating Graph Processing using
  {ReRAM}}. In \bibinfo{booktitle}{\emph{2018 IEEE International Symposium on
  High Performance Computer Architecture (HPCA)}}. IEEE,
  \bibinfo{pages}{531--543}.
\newblock


\bibitem[\protect\citeauthoryear{Srivastava, Jin, Liu, Albonesi, and
  Zhang}{Srivastava et~al\mbox{.}}{2020}]%
        {srivastava2020matraptor}
\bibfield{author}{\bibinfo{person}{Nitish Srivastava}, \bibinfo{person}{Hanchen
  Jin}, \bibinfo{person}{Jie Liu}, \bibinfo{person}{David Albonesi}, {and}
  \bibinfo{person}{Zhiru Zhang}.} \bibinfo{year}{2020}\natexlab{}.
\newblock \showarticletitle{{MatRaptor}: A Sparse-Sparse Matrix Multiplication
  Accelerator Based on Row-Wise Product}. In \bibinfo{booktitle}{\emph{2020
  53rd Annual IEEE/ACM International Symposium on Microarchitecture (MICRO)}}.
  IEEE, \bibinfo{pages}{766--780}.
\newblock


\bibitem[\protect\citeauthoryear{Swanson, Michelson, Schwerin, and
  Oskin}{Swanson et~al\mbox{.}}{2003}]%
        {swanson2003wavescalar}
\bibfield{author}{\bibinfo{person}{Steven Swanson}, \bibinfo{person}{Ken
  Michelson}, \bibinfo{person}{Andrew Schwerin}, {and} \bibinfo{person}{Mark
  Oskin}.} \bibinfo{year}{2003}\natexlab{}.
\newblock \showarticletitle{{WaveScalar}}. In
  \bibinfo{booktitle}{\emph{Proceedings of the 36th annual IEEE/ACM
  International Symposium on Microarchitecture}}. IEEE Computer Society,
  \bibinfo{pages}{291}.
\newblock


\bibitem[\protect\citeauthoryear{Taylor, Kim, Miller, Ghodrat, Greenwald,
  Johnson, Lee, Ma, Shnidman, Wentzlaff, et~al\mbox{.}}{Taylor
  et~al\mbox{.}}{2001}]%
        {taylor2001raw}
\bibfield{author}{\bibinfo{person}{Michael Taylor}, \bibinfo{person}{Jason
  Kim}, \bibinfo{person}{Jason Miller}, \bibinfo{person}{Fae Ghodrat},
  \bibinfo{person}{Ben Greenwald}, \bibinfo{person}{Paul Johnson},
  \bibinfo{person}{Walter Lee}, \bibinfo{person}{Albert Ma},
  \bibinfo{person}{Nathan Shnidman}, \bibinfo{person}{David Wentzlaff},
  {et~al\mbox{.}}} \bibinfo{year}{2001}\natexlab{}.
\newblock \showarticletitle{The {Raw} Processor: A Composeable 32-bit Fabric
  for Embedded and General Purpose Computing}. In
  \bibinfo{booktitle}{\emph{Proceedings of HotChips}},
  Vol.~\bibinfo{volume}{13}.
\newblock


\bibitem[\protect\citeauthoryear{Van~der Vorst}{Van~der Vorst}{1992}]%
        {van1992bi}
\bibfield{author}{\bibinfo{person}{Henk~A Van~der Vorst}.}
  \bibinfo{year}{1992}\natexlab{}.
\newblock \showarticletitle{{Bi-CGSTAB:} A Fast and Smoothly Converging Variant
  of {Bi-CG} for the Solution of Nonsymmetric Linear Systems}.
\newblock \bibinfo{journal}{\emph{SIAM Journal on scientific and Statistical
  Computing}} \bibinfo{volume}{13}, \bibinfo{number}{2} (\bibinfo{year}{1992}),
  \bibinfo{pages}{631--644}.
\newblock


\bibitem[\protect\citeauthoryear{Vilim, Rucker, Zhang, Liu, and Olukotun}{Vilim
  et~al\mbox{.}}{2020}]%
        {vilim2020gorgon}
\bibfield{author}{\bibinfo{person}{Matthew Vilim}, \bibinfo{person}{Alexander
  Rucker}, \bibinfo{person}{Yaqi Zhang}, \bibinfo{person}{Sophia Liu}, {and}
  \bibinfo{person}{Kunle Olukotun}.} \bibinfo{year}{2020}\natexlab{}.
\newblock \showarticletitle{Gorgon: Accelerating Machine Learning from
  Relational Data}. In \bibinfo{booktitle}{\emph{2020 ACM/IEEE 47th Annual
  International Symposium on Computer Architecture (ISCA)}}. IEEE,
  \bibinfo{pages}{309--321}.
\newblock


\bibitem[\protect\citeauthoryear{Vissers}{Vissers}{2019}]%
        {vissers2019versal}
\bibfield{author}{\bibinfo{person}{Kees Vissers}.}
  \bibinfo{year}{2019}\natexlab{}.
\newblock \showarticletitle{Versal: The {Xilinx} Adaptive Compute Acceleration
  Platform {(ACAP)}}. In \bibinfo{booktitle}{\emph{Proceedings of the 2019
  ACM/SIGDA International Symposium on Field-Programmable Gate Arrays}}.
  \bibinfo{pages}{83--83}.
\newblock


\bibitem[\protect\citeauthoryear{Wang, Davidson, Pan, Wu, Riffel, and
  Owens}{Wang et~al\mbox{.}}{2016}]%
        {wang2016gunrock}
\bibfield{author}{\bibinfo{person}{Yangzihao Wang}, \bibinfo{person}{Andrew
  Davidson}, \bibinfo{person}{Yuechao Pan}, \bibinfo{person}{Yuduo Wu},
  \bibinfo{person}{Andy Riffel}, {and} \bibinfo{person}{John~D Owens}.}
  \bibinfo{year}{2016}\natexlab{}.
\newblock \showarticletitle{Gunrock: A High-Performance Graph Processing
  Library on the {GPU}}. In \bibinfo{booktitle}{\emph{ACM SIGPLAN Notices}},
  Vol.~\bibinfo{volume}{51}. ACM, \bibinfo{pages}{11}.
\newblock


\bibitem[\protect\citeauthoryear{Wheeler}{Wheeler}{2020}]%
        {simplemachines}
\bibfield{author}{\bibinfo{person}{Bob Wheeler}.}
  \bibinfo{year}{2020}\natexlab{}.
\newblock \bibinfo{title}{Growing {AI} Diversity and Complexity Demands
  Flexible Data-Center Accelerators}.
\newblock
\newblock
\urldef\tempurl%
\url{https://www.simplemachines.ai/sites/default/files/SMI%20white%20paper-revised.pdf}
\showURL{%
\tempurl}


\bibitem[\protect\citeauthoryear{Wilson, Boe, Sala, Puttaswamy, and
  Zhao}{Wilson et~al\mbox{.}}{2009}]%
        {wilson2009user}
\bibfield{author}{\bibinfo{person}{Christo Wilson}, \bibinfo{person}{Bryce
  Boe}, \bibinfo{person}{Alessandra Sala}, \bibinfo{person}{Krishna~PN
  Puttaswamy}, {and} \bibinfo{person}{Ben~Y Zhao}.}
  \bibinfo{year}{2009}\natexlab{}.
\newblock \showarticletitle{User Interactions in Social Networks and their
  Implications}. In \bibinfo{booktitle}{\emph{Proceedings of the 4th ACM
  European conference on Computer systems}}. \bibinfo{pages}{205--218}.
\newblock


\bibitem[\protect\citeauthoryear{Zhang, Attaluri, Emer, and Sanchez}{Zhang
  et~al\mbox{.}}{2021a}]%
        {gamma}
\bibfield{author}{\bibinfo{person}{Guowei Zhang}, \bibinfo{person}{Nithya
  Attaluri}, \bibinfo{person}{Joel Emer}, {and} \bibinfo{person}{Daniel
  Sanchez}.} \bibinfo{year}{2021}\natexlab{a}.
\newblock \bibinfo{title}{Exploiting {Gustavson's} Algorithm to Accelerate
  Sparse Matrix Multiplication}.
\newblock
\newblock
\urldef\tempurl%
\url{https://asplos-conference.org/abstracts/asplos21-paper95-extended_abstract.pdf}
\showURL{%
\tempurl}


\bibitem[\protect\citeauthoryear{Zhang, Khoram, and Li}{Zhang
  et~al\mbox{.}}{2017}]%
        {zhang2017boosting}
\bibfield{author}{\bibinfo{person}{Jialiang Zhang}, \bibinfo{person}{Soroosh
  Khoram}, {and} \bibinfo{person}{Jing Li}.} \bibinfo{year}{2017}\natexlab{}.
\newblock \showarticletitle{Boosting the Performance of {FPGA}-Based Graph
  Processor using Hybrid Memory Cube: A Case for Breadth First Search}. In
  \bibinfo{booktitle}{\emph{Proceedings of the 2017 ACM/SIGDA International
  Symposium on Field-Programmable Gate Arrays}}. ACM,
  \bibinfo{pages}{207--216}.
\newblock


\bibitem[\protect\citeauthoryear{Zhang and Chen}{Zhang and Chen}{2018}]%
        {zhang2018gnn}
\bibfield{author}{\bibinfo{person}{Muhan Zhang} {and} \bibinfo{person}{Yixin
  Chen}.} \bibinfo{year}{2018}\natexlab{}.
\newblock \showarticletitle{Link Prediction Based on Graph Neural Networks}. In
  \bibinfo{booktitle}{\emph{Proceedings of the 32nd International Conference on
  Neural Information Processing Systems}} (Montr\'{e}al, Canada)
  \emph{(\bibinfo{series}{NIPS’18})}. \bibinfo{publisher}{Curran Associates
  Inc.}, \bibinfo{address}{Red Hook, NY, USA}, \bibinfo{pages}{5171–5181}.
\newblock


\bibitem[\protect\citeauthoryear{Zhang, Zhuo, Wang, Gao, Wu, Chen, Kozyrakis,
  and Qian}{Zhang et~al\mbox{.}}{2018b}]%
        {zhang2018graphp}
\bibfield{author}{\bibinfo{person}{Mingxing Zhang}, \bibinfo{person}{Youwei
  Zhuo}, \bibinfo{person}{Chao Wang}, \bibinfo{person}{Mingyu Gao},
  \bibinfo{person}{Yongwei Wu}, \bibinfo{person}{Kang Chen},
  \bibinfo{person}{Christos Kozyrakis}, {and} \bibinfo{person}{Xuehai Qian}.}
  \bibinfo{year}{2018}\natexlab{b}.
\newblock \showarticletitle{{GraphP}: Reducing Communication for PIM-Based
  Graph Processing with Efficient Data Partition}. In
  \bibinfo{booktitle}{\emph{2018 IEEE International Symposium on High
  Performance Computer Architecture (HPCA)}}. IEEE, \bibinfo{pages}{544--557}.
\newblock


\bibitem[\protect\citeauthoryear{Zhang, Du, Zhang, Lan, Liu, Li, Guo, Chen, and
  Chen}{Zhang et~al\mbox{.}}{2016}]%
        {zhang2016cambricon}
\bibfield{author}{\bibinfo{person}{Shijin Zhang}, \bibinfo{person}{Zidong Du},
  \bibinfo{person}{Lei Zhang}, \bibinfo{person}{Huiying Lan},
  \bibinfo{person}{Shaoli Liu}, \bibinfo{person}{Ling Li}, \bibinfo{person}{Qi
  Guo}, \bibinfo{person}{Tianshi Chen}, {and} \bibinfo{person}{Yunji Chen}.}
  \bibinfo{year}{2016}\natexlab{}.
\newblock \showarticletitle{{Cambricon-X}: An Accelerator for Sparse Neural
  Networks}. In \bibinfo{booktitle}{\emph{The 49th Annual IEEE/ACM
  International Symposium on Microarchitecture}}. IEEE Press,
  \bibinfo{pages}{20}.
\newblock


\bibitem[\protect\citeauthoryear{Zhang, Rucker, Vilim, Prabhakar, Hwang, and
  Olukotun}{Zhang et~al\mbox{.}}{2019}]%
        {zhang2019scalable}
\bibfield{author}{\bibinfo{person}{Yaqi Zhang}, \bibinfo{person}{Alexander
  Rucker}, \bibinfo{person}{Matthew Vilim}, \bibinfo{person}{Raghu Prabhakar},
  \bibinfo{person}{William Hwang}, {and} \bibinfo{person}{Kunle Olukotun}.}
  \bibinfo{year}{2019}\natexlab{}.
\newblock \showarticletitle{Scalable Interconnects for Reconfigurable Spatial
  Architectures}. In \bibinfo{booktitle}{\emph{Proceedings of the 46th
  International Symposium on Computer Architecture}}. ACM,
  \bibinfo{pages}{615--628}.
\newblock


\bibitem[\protect\citeauthoryear{Zhang, Yang, Baghdadi, Kamil, Shun, and
  Amarasinghe}{Zhang et~al\mbox{.}}{2018a}]%
        {zhang2018graphit}
\bibfield{author}{\bibinfo{person}{Yunming Zhang}, \bibinfo{person}{Mengjiao
  Yang}, \bibinfo{person}{Riyadh Baghdadi}, \bibinfo{person}{Shoaib Kamil},
  \bibinfo{person}{Julian Shun}, {and} \bibinfo{person}{Saman Amarasinghe}.}
  \bibinfo{year}{2018}\natexlab{a}.
\newblock \showarticletitle{{GraphIt}: A High-Performance Graph {DSL}}.
\newblock \bibinfo{journal}{\emph{Proc. ACM Program. Lang.}}
  \bibinfo{volume}{2}, \bibinfo{number}{OOPSLA}, Article
  \bibinfo{articleno}{121} (\bibinfo{date}{Oct.} \bibinfo{year}{2018}),
  \bibinfo{numpages}{30}~pages.
\newblock
\urldef\tempurl%
\url{https://doi.org/10.1145/3276491}
\showDOI{\tempurl}


\bibitem[\protect\citeauthoryear{Zhang, Zhang, Zhao, Vilim, Shahbaz, and
  Olukotun}{Zhang et~al\mbox{.}}{2021b}]%
        {zhang2021sara}
\bibfield{author}{\bibinfo{person}{Yaqi Zhang}, \bibinfo{person}{Nathan Zhang},
  \bibinfo{person}{Tian Zhao}, \bibinfo{person}{Matt Vilim},
  \bibinfo{person}{Muhammad Shahbaz}, {and} \bibinfo{person}{Kunle Olukotun}.}
  \bibinfo{year}{2021}\natexlab{b}.
\newblock \showarticletitle{SARA: Scaling a Reconfigurable Dataflow
  Accelerator}. In \bibinfo{booktitle}{\emph{2021 ACM/IEEE 48th Annual
  International Symposium on Computer Architecture (ISCA)}}. IEEE,
  \bibinfo{pages}{1041--1054}.
\newblock


\bibitem[\protect\citeauthoryear{Zhang, Wang, Han, and Dally}{Zhang
  et~al\mbox{.}}{2020}]%
        {zhang2020sparch}
\bibfield{author}{\bibinfo{person}{Zhekai Zhang}, \bibinfo{person}{Hanrui
  Wang}, \bibinfo{person}{Song Han}, {and} \bibinfo{person}{William~J Dally}.}
  \bibinfo{year}{2020}\natexlab{}.
\newblock \showarticletitle{{SpArch}: Efficient Architecture for Sparse Matrix
  Multiplication}. In \bibinfo{booktitle}{\emph{2020 IEEE International
  Symposium on High Performance Computer Architecture (HPCA)}}. IEEE,
  \bibinfo{pages}{261--274}.
\newblock


\bibitem[\protect\citeauthoryear{Zhao, Zhang, and Olukotun}{Zhao
  et~al\mbox{.}}{2019}]%
        {zhao2019serving}
\bibfield{author}{\bibinfo{person}{Tian Zhao}, \bibinfo{person}{Yaqi Zhang},
  {and} \bibinfo{person}{Kunle Olukotun}.} \bibinfo{year}{2019}\natexlab{}.
\newblock \showarticletitle{Serving Recurrent Neural Networks Efficiently with
  a Spatial Accelerator}.
\newblock \bibinfo{journal}{\emph{arXiv preprint arXiv:1909.13654}}
  (\bibinfo{year}{2019}).
\newblock


\bibitem[\protect\citeauthoryear{Zhou, Chelmis, and Prasanna}{Zhou
  et~al\mbox{.}}{2015}]%
        {zhou2015accelerating}
\bibfield{author}{\bibinfo{person}{Shijie Zhou}, \bibinfo{person}{Charalampos
  Chelmis}, {and} \bibinfo{person}{Viktor~K Prasanna}.}
  \bibinfo{year}{2015}\natexlab{}.
\newblock \showarticletitle{Accelerating Large-Scale Single-Source Shortest
  Path on {FPGA}}. In \bibinfo{booktitle}{\emph{2015 IEEE International
  Parallel and Distributed Processing Symposium Workshop}}. IEEE,
  \bibinfo{pages}{129--136}.
\newblock


\end{thebibliography}

\end{document}